\documentclass[11pt,letterpaper]{article}
\pdfoutput=1
\usepackage{jheppub} 
\usepackage{mathrsfs}
\usepackage{slashed}
\usepackage{caption}
\usepackage{epstopdf}
\usepackage[normalem]{ulem}
\usepackage[bottom]{footmisc}
\usepackage{subcaption}
\usepackage{bbold}
\usepackage{titlesec}
\usepackage{threeparttable}
\usepackage{booktabs} 
\usepackage{changepage}
\usepackage[utf8]{inputenc}
\usepackage{dsfont}
\usepackage{grffile}
\usepackage{graphicx}  
\usepackage{dcolumn}   
\usepackage{bm}        
\usepackage{amssymb}   
\usepackage{setspace}
\usepackage{amsmath, amssymb, setspace}
\usepackage{array}
\usepackage{booktabs}
\usepackage{caption}
\usepackage{indentfirst}
\usepackage{float}
\usepackage{lmodern}
\usepackage{multirow}
\usepackage{soul}
\usepackage[normalem]{ulem}
\usepackage{braket}
\usepackage{comment}
\usepackage{appendix}  
\usepackage{feynmp-auto} 
\graphicspath{{figs/}}
\usepackage{enumitem}

\title{\huge{Unified Gas Heating Constraints on Extended Dark Matter Compact Objects}}

\author[a]{TaeHun Kim,}
\author[a]{Philip Lu,}
\author[b,c,d,e]{Volodymyr Takhistov}

\affiliation[a]{School of Physics, Korea Institute for Advanced Study,
85 Hoegi-ro, Dongdaemun-gu, Seoul 02455, Republic of Korea
}
\affiliation[b]{International Center for Quantum-field Measurement Systems for Studies of the Universe and Particles (QUP, WPI),
High Energy Accelerator Research Organization (KEK), \\ Oho 1-1, Tsukuba, Ibaraki 305-0801, Japan}
\affiliation[c]{Theory Center, Institute of Particle and Nuclear Studies (IPNS), High Energy Accelerator Research Organization (KEK), Tsukuba 305-0801, Japan}
\affiliation[d]{Kavli Institute for the Physics and Mathematics of the Universe (WPI), \\ The University of Tokyo Institutes for Advanced Study, The University of Tokyo, \\ Kashiwa, Chiba 277-8583, Japan}
\affiliation[e]{Graduate University for Advanced Studies (SOKENDAI),  
1-1 Oho, Tsukuba, Ibaraki 305-0801, Japan}  

\emailAdd{gimthcha@kias.re.kr}
\emailAdd{philiplu11@gmail.com}
\emailAdd{vtakhist@post.kek.jp} 

\abstract{
We present the first unified constraints on a broad class of extended dark matter compact objects (EDCOs) from interstellar gas heating. These include axion stars, Q-balls, axion miniclusters, dark fermion stars and primordial black holes surrounded by dark matter halos, which arise in a wide range of theories beyond the Standard Model. As such massive objects traverse the interstellar medium, their gravitational influence generates wakes and, if sufficiently compact, drives accretion flows that heat gas in their vicinity. Our general framework extends standard dynamical friction treatments by incorporating finite-size effects, internal density profiles, gas penetration through objects, and criteria for accretion disk formation. We perform detailed numerical calculations of wake formation and gas heating and apply our results to the Leo T dwarf galaxy, establishing new constraints on the dark matter fraction in EDCOs heavier than a solar mass spanning several orders of magnitude in both mass and abundance.
}

\allowdisplaybreaks

\begin{document}

\preprint{KEK-QUP-2025-0018, KEK-TH-2748}
 \maketitle
\flushbottom
 
\section{Introduction}

The nature of dark matter (DM), which constitutes approximately 85\% of the total matter content of the Universe, remains one of the most significant open questions in physics. While its gravitational influence on galactic and cosmological scales is firmly established, possible non-gravitational interactions of DM and its small-scale structure remain unknown. Extensive experimental and observational programs have been dedicated to probing particle-like DM candidates, while gravitational and astrophysical observations have been employed to constrain massive compact DM objects—such as primordial black holes (PBHs)—that could have formed in the early Universe (see e.g.~\cite{Bertone:2004pz,Gelmini:2015zpa} for review).

In a variety of theories beyond the Standard Model (SM), macroscopic self-gravitating configurations composed of dark sector fields arise and such extended DM compact objects (EDCOs) can comprise a sizable fraction of the total DM abundance.  Examples include axion stars~\cite{Schive:2014dra,Schive:2014hza,Chavanis:2011zi,Chavanis:2011zm,Chavanis:2016dab,Visinelli:2017ooc,Chang:2024fol}, Q-balls~\cite{Coleman:1985ki,Ansari:2023cay,Heeck:2020bau}, axion minicluster subhalos
~\cite{Xiao:2021nkb,Chang:2024fol}  and fermionic dark stars~\cite{Narain:2006kx,Gresham:2018rqo}. In addition, PBHs constituting only a subdominant fraction of DM as suggested by existing constraints over a wide range of parameter space can be generically surrounded by massive DM halos, forming dressed PBHs (dPBHs)~\cite{Mack:2006gz,Berezinsky:2013fxa,Adamek:2019gns,Boudaud:2021irr,Oguri:2022fir,GilChoi:2023ahp}. Compared to nearly point-like PBHs or highly diffuse subhalos, EDCOs have finite size and internal structure and, in some cases, can permit interstellar gas composed of SM particles to pass through them, leading to qualitative differences in their gravitational interactions with astrophysical media.

In this work, we put forth a general probe of EDCOs based on their interactions with interstellar and galactic gas. As EDCOs traverse gas-rich environments, they induce dynamical friction wakes and, when sufficiently compact, can sustain accretion flows. The resulting energy dissipation can heat the surrounding cold gas thereby altering its thermal state and leaving observable imprints in a variety of astrophysical systems. Interactions of particle DM with astrophysical gas and galactic media have been extensively studied~\cite{Chivukula:1989cc,Dubovsky:2015cca,Bhoonah:2018wmw,Wadekar:2019xnf}. In Refs.~\cite{Lu:2020bmd,Takhistov:2021aqx,Kim:2020ngi,Laha:2020vhg,Takhistov:2021upb}, it was demonstrated that heating of astrophysical gas can place strong constraints on PBHs as compact macroscopic DM candidates.

We present a general framework to analyze dynamical friction and gas-heating rates for EDCOs that includes finite-radius effects, realistic internal density profiles and penetration of gas through the objects—features absent from standard point-like mass treatments such as for PBHs. While Ref.~\cite{Wadekar:2022ymq} considered extended DM objects following earlier PBH analyses~\cite{Lu:2020bmd,Takhistov:2021aqx,Kim:2020ngi,Laha:2020vhg,Takhistov:2021upb}, their approach employed the same hard-sphere point-mass formalism for dynamical friction with an effective size cutoff. In contrast, our treatment accounts for finite-size effects through direct integration over realistic mass profiles and models extended wake generation inside compact objects that are transparent to SM particles.
We further connect mass–radius relations to the underlying microphysics of each EDCO. 
In addition, we modify the accretion disk formalism to incorporate varying mass–radius relations and delineate the regimes in which accretion flows develop around sufficiently compact and massive objects.

Applying our framework to systems with well-measured gas densities and temperatures, notably Leo T, we derive constraints on the abundance of extended compact DM structures across several decades in mass and fractional contribution, from stellar to supermassive scales. These bounds are distinct from and complementary to other probes of EDCO, such as lensing~(e.g.~\cite{Croon:2020wpr,DeRocco:2023hij}).
We stress that our analysis is conceptually distinct from constraints based on heating of a ``gas'' composed of stars~\cite{Graham:2023unf,Graham:2024hah}, which concern increases in stellar velocity dispersion rather than heating of interstellar gas.
Our results open a new avenue for testing compact dark sector configurations through their dynamical impact on ordinary matter.

The study is organized as follows. In Sec.~\ref{sec:gasheating}, we give an overview of the gas heating formalism due to EDCOs and describe the Leo T dwarf galaxy system. Sec.~\ref{sec:dynfric} develops the analysis of dynamical friction for EDCOs, including a numerical treatment of wake formation that goes beyond the standard approach. Sec.~\ref{sec:disk} presents the analysis of accretion disk formation by EDCOs. In Sec.~\ref{sec:selected}, we examine gas heating from a range of EDCOs including axion stars, axion miniclusters, Q-balls, dark fermion stars, and PBHs dressed with dark halos based on existing microphysical models  and derive  the corresponding new constraints for each case. We conclude in Sec.~\ref{sec:conclusions}.

\section{Gas Heating by Macroscopic Dark Matter
}
\label{sec:gasheating}

We review the formalism for analyzing gas heating in astrophysical systems, as previously applied to constrain PBHs in Refs.~\cite{Lu:2020bmd,Takhistov:2021aqx,Takhistov:2021upb}. The gas heating approach considers the impact of compact objects on the thermal equilibrium of a stable gas system or cloud, which can be assumed to maintain an approximately constant temperature over a timescale comparable to its cooling time, typically exceeding millions of years. This criterion disfavors gas clouds that can be dynamically disrupted by high-velocity outflows or other destabilizing processes~\cite{Farrar:2019qrv}. Among the viable gas systems identified in Ref.~\cite{Wadekar:2019xnf}, Ref.~\cite{Takhistov:2021aqx} found the dwarf galaxy Leo~T to be particularly well suited for analyzing PBH effects on surrounding gas. Leo T likewise constitutes an ideal target for constraining EDCOs, as it is gas-rich while being DM dominated, contains a stable reservoir of cold gas with low gas–DM velocity dispersion, experiences minimal stellar heating, and has well-measured astrophysical properties. Throughout, we work in natural units $\hbar=c=k_B=1$, except in expressions for accretion disk emission where we restore constants for clarity.

The volumetric cooling rate $\dot{C}$ can be compared to volumetric heating rate 
$H(M) \rho_{\rm DM}/M $, where $H(M)$ is the total heating per compact object of mass $M$ and $\rho_{\rm DM}$ is the local DM density in the gas system. Imposing thermal equilibrium then yields an upper bound on the allowed DM fraction in EDCOs 
\begin{equation}
f_{\rm DM}(M) < f_{\rm bound}(M) = \max \left[ \frac{M   \dot{C}}{\rho_{\rm DM}   H(M)},   \frac{3M}{4\pi r_{\rm sys}^3 \rho_{\rm DM}} \right] ,
\label{eq:fbound}
\end{equation}
where the second term on the right corresponds to the ``incredulity’’ limit requiring that at least one EDCO exists within a system of size $r_{\rm sys}$. Since $f_{\rm bound}$ scales inversely with $H(M)$, the constraints become more stringent for EDCOs with more efficient heating deposition mechanisms.

The heating processes we consider—dynamical friction, accretion disk photon emission and outflows\footnote{Here we do not consider the formation of jets, which can efficiently occur for spinning PBHs~\cite{Takhistov:2021upb}.}—all depend on velocities of EDCOs. The relative velocity distribution between a compact DM object  and the gas cloud is given by~\cite{Takhistov:2021aqx}
\begin{equation}
\dfrac{d f_v}{dv} = \dfrac{v}{\sqrt{2\pi} \sigma_v   v_b} \left[ e^{-(v-v_b)^2 / (2\sigma_v^2)} - e^{-(v+v_b)^2 / (2\sigma_v^2)} \right] ,
\label{eq:vdist}
\end{equation}
where $v_b$ is the bulk velocity, such as rotational motion in the Milky Way, between the gas system and the compact objects, and $\sigma_v$ is the relative velocity dispersion. In the absence of bulk motion with $v_b \simeq 0$, as in Leo~T, this reduces to the well known Maxwell–Boltzmann form
\begin{equation}
\dfrac{d f_v}{dv} \simeq \sqrt{\dfrac{2}{\pi}}   \dfrac{v^2}{\sigma_v^3}   e^{-v^2 / (2\sigma_v^2)} . \label{eq:MB}
\end{equation}
The velocity-dependent heating rate $\mathcal{H}(M,v)$ is averaged over the velocity distribution to obtain the total heating rate per EDCO 
\begin{equation}
H(M) = \int_{0}^{\infty} dv  \dfrac{d f_v}{dv}   \mathcal{D}   \mathcal{H}(M,v) , 
\label{eq:HM}
\end{equation}
where the additional factor $\mathcal{D}$ captures efficiency that can be time-dependent  
for outflow duty cycles as discussed Sec.~\ref{ssec:addfactors}. Contributions from dynamical friction, photon emission and outflows are included in $\mathcal{H}(M,v)$. The resulting $H(M)$ is then considered in Eq.~\eqref{eq:fbound} to evaluate the allowed EDCO DM fraction $f_{\rm DM}$. 

Throughout we neglect possible additional heating from astrophysical and stellar sources~\cite{Wadekar:2022ymq}, focusing instead on heating that arises directly and indirectly from gravitational forces due to EDCOs. We also do not consider possible particle DM–gas interactions involving the compact objects, leaving their analysis for future work. This results in our bounds being conservative, since any additional heating sources would only strengthen the constraints.

\subsection{Target system Leo T} \label{ssec:LeoT}

As our primary target system for analyzing the effects of EDCOs, we focus on the central gas region of the Leo T dwarf galaxy. Several factors position Leo T as a particularly favorable system for considering gas heating by macroscopic objects. The estimated DM velocity dispersion is low with $\sigma_v \simeq 6.9~\mathrm{km~s^{-1}}$ and the sound speed is $c_s \simeq 9~\mathrm{km~s^{-1}}$~\cite{Faerman:2013pmm,Ryan-Weber:2007guk}. The corresponding Mach numbers, $\mathcal{M} = v/c_s \sim \mathcal{O}(1)$, are near-optimal for maximizing dynamical friction.
Further, these parameters lead to high Bondi–Hoyle accretion rates as discussed below in~Eq.~\eqref{eq:bondihoyle}.

We consider the Leo T's central region of radius $r_{\rm sys} = 350~\mathrm{pc}$ where the hydrogen gas is predominantly neutral, with an average density $n \simeq 0.07~\mathrm{cm^{-3}}$~\cite{Faerman:2013pmm,Ryan-Weber:2007guk} and metallicity $[\mathrm{Fe/H}] \simeq -2$~\cite{Kirby:2008ab}. The low metallicity yields a correspondingly low gas cooling rate of $\dot{C} = 2.28 \times 10^{-30}~\mathrm{erg~cm^{-3}~s^{-1}}$~\cite{Wadekar:2019xnf,Kim:2020ngi}, limiting the amount of heating that can be radiatively dissipated. Further, the DM density in this region is $\rho_{\rm DM} \simeq 1.75~\mathrm{GeV~cm^{-3}}$, giving a total enclosed DM mass of $(4\pi/3) \rho_{\rm DM} r_{\rm sys}^3 \simeq 10^7~M_\odot$. Hence, the requirement that at least one object resides within the system imposes a lower bound on the possible constraints for the EDCO DM fraction
\begin{equation}
f_{\rm bound} \gtrsim \dfrac{M}{  10^7~M_\odot }~. 
\end{equation}

Variety of other possible target astrophysical systems can be analyzed using the same framework, such as gas clouds of Milky Way~\cite{Wadekar:2019xnf,Takhistov:2021aqx,Wadekar:2022ymq}. In the case of Milky Way gas clouds, the 
sensitivity to DM fraction 
contributing to gas heating can be expected to be orders of magnitude weaker than for Leo~T. The higher metallicity of these atomic hydrogen clouds increases their cooling rate, which is determined by their trace oxygen and carbon content~\cite{Kim:2020ngi}. Furthermore, the large bulk rotational velocity $v_b \simeq 220~\mathrm{km~s^{-1}}$ and DM velocity dispersion $\sigma_v \simeq 124~\mathrm{km~s^{-1}}$ in the Milky Way significantly suppresses the Bondi–Hoyle accretion rates. More so, the total DM mass in each of these clouds is only $\mathcal{O}(10^3)~M_\odot$, limiting the range of compact object masses that can be probed~\cite{Takhistov:2021aqx}.

\section{Dynamical Friction Heating}
\label{sec:dynfric}

The gravitational influence of a massive object traversing a medium generates a density wake. In turn, the perturbed wake exerts a reaction force that slows the object resulting in dynamical friction or gravitational drag. This self-friction is a general mechanism by which EDCOs can transfer energy to, and heat, the surrounding medium purely through gravitational interaction. We begin by summarizing the standard point-mass formalism, which models dynamical friction as arising from a Dirac delta–function mass distribution with an impenetrable boundary. We then develop a new framework for extended mass distributions with transparent boundaries, in which the gravitational drag is obtained by integrating the contributions from each infinitesimal mass element of the object.

\subsection{Point-mass formalism}

To discuss dynamical friction of EDCOs we first review an idealized case of a point mass in an infinite medium, as analytically described in Ref.~\cite{Ostriker:1998fa}.
Consider a point mass $M_{\rm obj}$ moving in a uniform gaseous medium of average density $\bar{\rho}_{\rm g}$ with velocity $\vec{v} = \mathcal{M} c_s \hat{z}$ in direction $\hat{z}$, where $c_s$ is the sound speed and $\mathcal{M}$ is the Mach number. The object’s gravitational potential $\Phi_{\rm obj}$ induces a density perturbation $\alpha$, defined by
\begin{equation}
\rho(\vec{x}, t) = \bar{\rho}_{\rm g} \left[ 1 + \alpha(\vec{x}, t) \right] ,
\end{equation}
as a function of perturbation coordinate $\vec{x}$ and time $t$.

In the linear regime, $\alpha$ satisfies the inhomogeneous wave equation
\begin{equation}
\vec{\nabla}^2 \alpha - \frac{1}{c_s^2} \frac{\partial^2 \alpha}{\partial t^2} = - \frac{1}{c_s^2} \vec{\nabla}^2 \Phi_{\rm obj}~.
\label{eq:alphawaveeq}
\end{equation}
The gravitational potential $\Phi_{\rm obj}$ obeys Poisson’s equation 
\begin{equation}
\vec{\nabla}^2 \Phi_{\rm obj} = 4\pi G \rho_{\rm obj}~,
\end{equation}
where $\rho_{\rm obj}$ is the object’s mass density and $G$ is the gravitational constant.
Using this relation, the solution for $\alpha$ can be expressed via the retarded Green’s function of Eq.~\eqref{eq:alphawaveeq} as
\begin{equation}
    \alpha(\vec{x}, t) = \frac{G}{c_s^2}\iint d^3 \vec{x}' dt' \, \frac{\delta(t' - (t - |\vec{x} - \vec{x}'|/c_s))}{|\vec{x} - \vec{x}'|} \, \rho_{\rm obj} (\vec{x}', t')~. \label{eq:greenfuncint}
\end{equation} 
Hence, the density perturbation wake $ \alpha(\vec{x}, t)$ is sourced by the object’s mass distribution at earlier times with the signal propagating at the sound speed $c_s$. Here, Eq.~\eqref{eq:greenfuncint} describes the  response of the medium to a moving gravitational source that gives rise to the characteristic wake structure and Mach cone for supersonic motion.

\begin{figure}[t]
\centering
\includegraphics[width=0.49\linewidth]{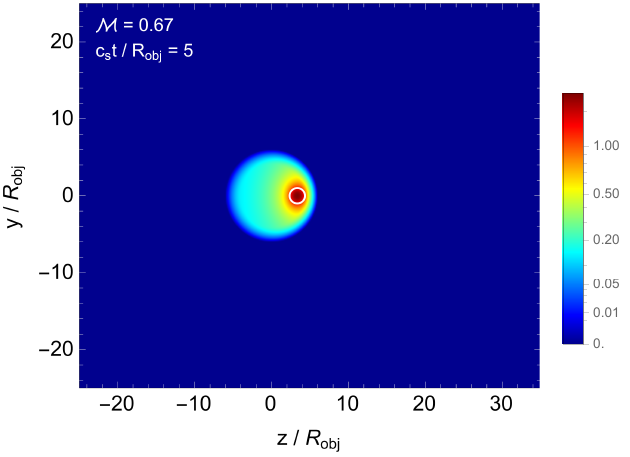}
\includegraphics[width=0.49\linewidth]{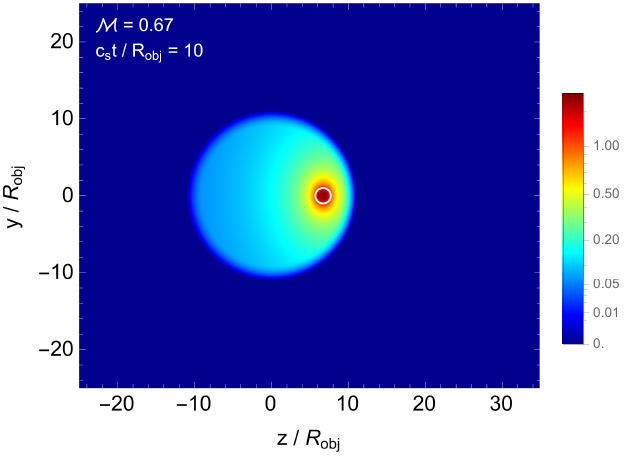}\\
\includegraphics[width=0.49\linewidth]{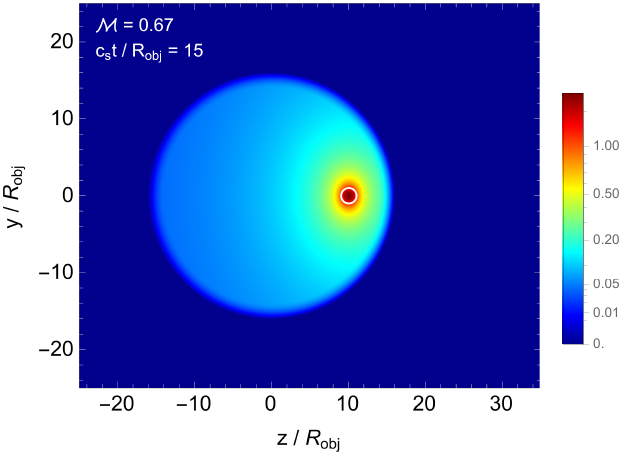}
\includegraphics[width=0.49\linewidth]{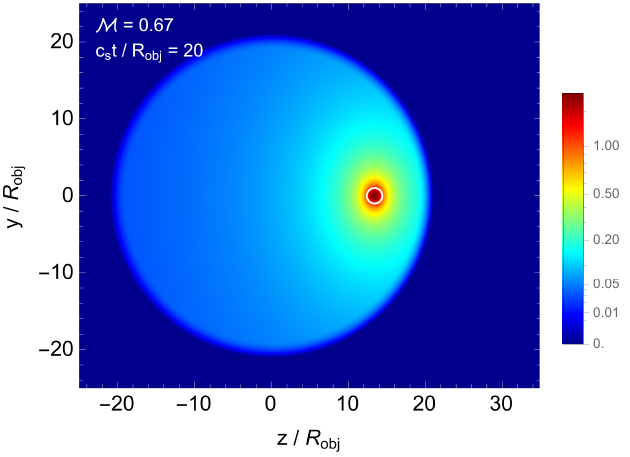}
\caption{Density wake $\alpha(\vec{x},t)$ generated by a spherically symmetric EDCO (white circle) with a uniform density profile moving in direction $\hat{z}$ subsonically with $\mathcal{M} = 0.67$ through a uniform gaseous medium, for  $c_s t / R_{\rm obj}$ = 5 [Top left], 10 [Top right], 15 [Bottom left], 20 [Bottom right]. 
The perturbation forms a sonic sphere centered at the origin and peaked around the object. 
No Mach cone is present in the subsonic regime. The wake extends inside EDCO, in contrast to the point mass case, producing a smoother density profile and affecting Coulomb logarithm.
}
\label{fig:wakesubsonic}
\end{figure}

\begin{figure}[t]
\centering
\includegraphics[width=0.49\linewidth]{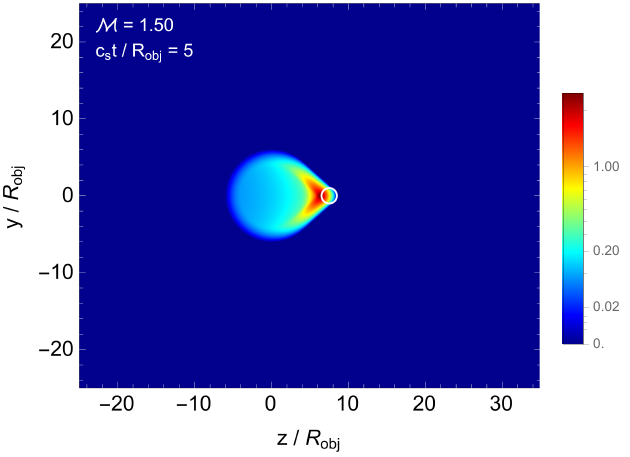}
\includegraphics[width=0.49\linewidth]{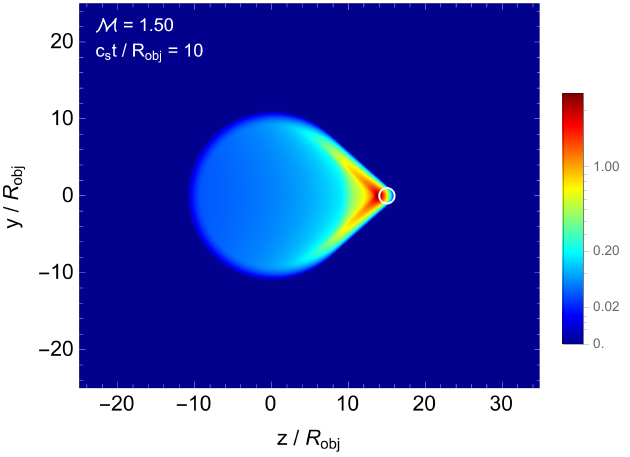}\\
\includegraphics[width=0.49\linewidth]{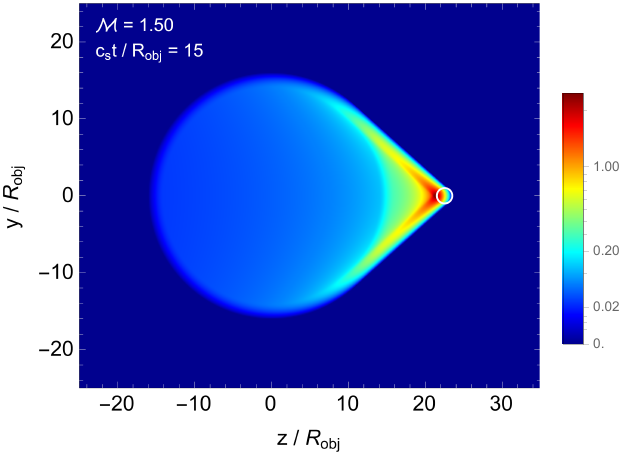}
\includegraphics[width=0.49\linewidth]{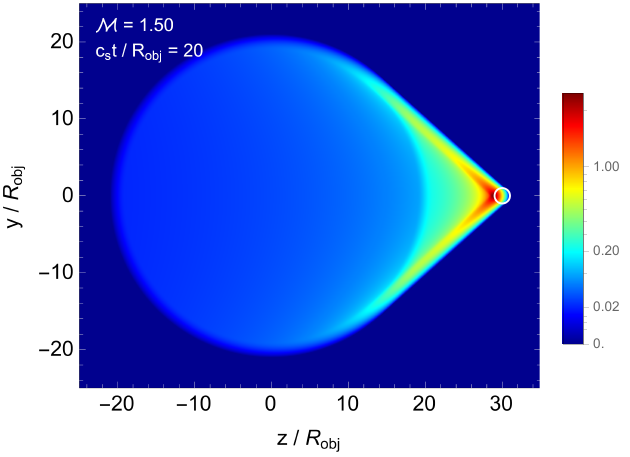}
\caption{Density wake $\alpha(\vec{x},t)$ generated by a spherically symmetric EDCO (white circle) with a uniform density profile moving in direction $\hat{z}$ supersonically with $\mathcal{M} = 1.50$ through a uniform gaseous medium, for  $c_s t / R_{\rm obj}$ = 5 [Top left], 10 [Top right], 15 [Bottom left], 20 [Bottom right]. The perturbations consist of a sonic sphere and a pronounced Mach cone trailing the object, within which wavefronts from successive positions of the object overlap. The wake amplitude is enhanced inside the cone, as reflected by the factor of two in $\mathcal{S}(\vec{x},t)$ for $\mathcal{M} > 1$ in Eq.~\eqref{eq:Spointmass}. The wake extends inside EDCO, in contrast to the point mass case, producing a smoother density profile and affecting Coulomb logarithm. 
} 
\label{fig:wakesupersonic}
\end{figure}

For a point mass, evaluating Eq.~\eqref{eq:greenfuncint} with a Dirac delta function density profile $\rho_{\rm obj}$ yields the wake
\begin{equation}
    \alpha(\vec{x}, t) = \frac{G M_{\rm obj}   \mathcal{S}(\vec{x}, t)}{c_s^2 [(x^2 + y^2)(1-\mathcal{M}^2) + (z - \mathcal{M} c_s t)^2]^{1/2}}~, \label{eq:alphapointmass}
\end{equation}
where the coordinates consider the object is at the origin at $t = 0$.
The density perturbation consists of a sonic sphere propagating from the origin for all Mach numbers, and when the motion is supersonic an additional Mach cone trailing behind the object appears, so that 
\begin{equation}
    \mathcal{S}(\vec{x}, t) = 
    \begin{cases}
        1 & \text{if } |\vec{x}| < c_s t \quad \text{for all } \mathcal{M}   , \\[6pt]  
        2 & \text{if } |\vec{x}| > c_s t, \ \ \dfrac{c_s t}{\mathcal{M}} < z < \mathcal{M} c_s t, \ 
        (\mathcal{M}^2 - 1)(x^2 + y^2) < (\mathcal{M} c_s t - z)^2  \\ 
          &     \text{for } \mathcal{M} > 1   , \\[6pt]
        0 & \text{otherwise.}
    \end{cases}
    \label{eq:Spointmass}
\end{equation}

In Fig.~\ref{fig:wakesubsonic} and Fig.~\ref{fig:wakesupersonic}, we illustrate numerically evaluated spatial structure of the density wake $\alpha$ given by Eq.~\eqref{eq:alphapointmass} for subsonic and supersonic motion, respectively.  At each moment, the disturbance propagates isotropically at the sound speed, producing a spherical wavefront (i.e. sonic sphere) centered on the object’s instantaneous position.
For $\mathcal{M} < 1$ in Fig.~\ref{fig:wakesubsonic}, the object moves slower than the sound speed and thus the entire disturbance produced at $t \geq 0$ remains confined within the spherical wavefront that originated at $t = 0$ from the object’s initial position at the origin.
For $\mathcal{M} > 1$ in   Fig.~\ref{fig:wakesupersonic}, the motion of the object outpaces the sound waves it generates, resulting in the formation of a Mach cone whose opening angle satisfies $\sin\theta_{\rm M} = 1/\mathcal{M}$. Within the cone, wavefronts overlap coherently leading to an enhanced density contrast.

The gravitational ``drag'' force associated with dynamical friction, exerted by the wake on the moving object, is given by
\begin{equation}
    F_{\rm DF} = \int d^3 x \, \frac{G M_{\rm obj} \, \bar{\rho}_g \alpha (\vec{x}, t)   (z - \mathcal{M} c_s t)}{[x^2 + y^2 + (z - \mathcal{M} c_s t)^2]^{3/2}}~. \label{eq:FDFpointmassintegral}
\end{equation}
It is convenient to separate the force into a prefactor depending on $M_{\rm obj}$ and $\bar{\rho}_g$, and a purely geometric factor $I(\mathcal{M},t)$ that depends only on Mach number and time
\begin{equation}
    F_{\rm DF} (t) = \frac{4\pi (G M_{\rm obj})^2 \bar{\rho}_g}{(\mathcal{M} c_s)^2}   I(\mathcal{M}, t)~.
    \label{eq:FDFparameterize}
\end{equation}
For a point mass, inserting Eq.~\eqref{eq:alphapointmass} and Eq.~\eqref{eq:Spointmass} into Eq.~\eqref{eq:FDFpointmassintegral} yields
\begin{equation}
    I(\mathcal{M}, t) = I_{\rm pm}(\mathcal{M}, t) = 
    \begin{cases}
		\dfrac{1}{2} \ln \left( \dfrac{1+\mathcal{M}}{1-\mathcal{M}} \right) - \mathcal{M} & \text{for} \quad \mathcal{M} < 1 ,\\ \\
        \dfrac{1}{2} \ln \left(1 - \dfrac{1}{\mathcal{M}^2} \right) + \ln \dfrac{\mathcal{M} c_s t}{r_{\rm min}} & \text{for} \quad \mathcal{M} > 1 ,
	\end{cases}\label{eq:IMachpointmass}
\end{equation}
where $r_{\rm min}$ is the minimum separation between the object and the medium that regulates the lower integration limit and $\mathcal{M} c_s t$ provides the upper cut-off bounded by the finite size of the system $r_{\rm sys}$.
The kinetic energy lost by the object through dynamical friction is transferred to the medium at the heating rate
\begin{equation}
\mathcal{H}(M,v) = F_{\rm DF}  \mathcal{M} c_s~.
\end{equation}

Thus far, the discussion has considered the point-mass treatment of objects generating wakes. In Ref. \cite{Wadekar:2022ymq} finite-size effects were modeled by simply adjusting the effective $r_{\rm min}$ in the point-mass Coulomb logarithm of Eq.~\eqref{eq:IMachpointmass}. In contrast, as discussed below, we perform a full numerical evaluation of dynamical friction for extended objects by integrating the wake formation over their complete mass distribution, explicitly allowing gas composed of SM constituents to penetrate the EDCO interior and generate internal wakes.

\subsection{Finite-size effects}

Physically, the choice of the minimum cutoff scale $r_{\rm min}$ depends on the nature of the compact object and the medium. For black holes in a gaseous environment, the natural cutoff is the Schwarzschild radius $R_s = 2 G M_{\rm obj}$.
Although the linear approximation of Eq.~\eqref{eq:alphawaveeq} breaks down before reaching this scale, general relativistic corrections near the horizon are typically negligible compared to the large logarithmic contribution in the force. In contrast, for consideration of objects interacting with a ``gas'' of stars, the cutoff can be set by the impact parameter of stellar encounters~\cite{Graham:2023unf,Graham:2024hah}. For extended solid objects such as neutron stars, $r_{\rm min}$ is naturally given by the physical radius of the object  $R_{\rm obj}$. However, unlike the purely gravitational case, such bodies can also disturb the medium through direct hard-sphere scattering collisions, so the wake structure deviates from Eq.~\eqref{eq:alphawaveeq} and Eq.~\eqref{eq:alphapointmass}, and the total drag includes additional non-gravitational contributions.

For EDCOs that are transparent to gas composed of SM particles, the medium is perturbed only through gravity. The wake can then be obtained by integrating the contributions from each infinitesimal mass element over the object’s volume, thereby modifying Eq.~\eqref{eq:alphapointmass}. Similarly, the dynamical friction force in Eq.~\eqref{eq:FDFpointmassintegral} generalizes to a double integral over both the wake profile and the object’s internal mass distribution. We analyze a variety of EDCOs under this assumption in Sec.~\ref{sec:selected}. Possible non-gravitational interactions between EDCO dark-sector constituents and Standard Model particles (e.g. Ref.~\cite{Wadekar:2022ymq}) could alter the wake structure and render the objects partially opaque. A detailed analysis of such effects is beyond the scope of this work and is left for future study.

We now analyze the dynamical friction of an EDCO with radius $R_{\rm obj}$ and a spherically symmetric density profile $\rho_{\rm obj}(r)$. The wake is obtained by integrating Eq.~\eqref{eq:alphapointmass} over the contributions from each mass element at an internal coordinate $\vec{x}_{\rm obj}$, defined with respect to the object’s center, so that $|\vec{x}_{\rm obj}| \leq R_{\rm obj}$. This yields
\begin{equation}
    \alpha(\vec{x}, t) = \int d^3 x_{\rm obj} \, \frac{G \rho_{\rm obj}(\vec{x}_{\rm obj})    \mathcal{S}(\vec{x}, t; \vec{x}_{\rm obj})}{c_s^2 \{[(x-x_{\rm obj})^2 +(y-y_{\rm obj})^2] (1- \mathcal{M}^2) + (z-z_{\rm obj} - \mathcal{M} c_s t)^2\}^{1/2}}  \label{eq:alphaextobj}
\end{equation}
where $\mathcal{S}$ now depends explicitly on the internal coordinate
\begin{equation}
\label{eq:S}
    \mathcal{S}(\vec{x}, t; \vec{x}_{\rm obj}) = 
    \begin{cases}
        1 & \text{if} \quad |\vec{x} - \vec{x}_{\rm obj}| < c_s t \quad \text{for all} \quad  \mathcal{M} , \\  
        2 & \text{if} \quad |\vec{x} - \vec{x}_{\rm obj}| > c_s t  , \ \ \dfrac{c_s t}{\mathcal{M}} < z - z_{\rm obj} < \mathcal{M} c_s t\\ &   (\mathcal{M}^2 - 1)[(x-x_{\rm obj})^2 + (y-y_{\rm obj})^2] < [\mathcal{M} c_s t  - (z-z_{\rm obj})]^2  \\ &  \text{for} \quad \mathcal{M} > 1 ,  \\
        0 & \text{otherwise} .
    \end{cases}
\end{equation}

The dynamical friction force then generalizes to a double integral over both the wake and the object’s internal mass distribution
\begin{eqnarray}
    F_{\rm DF} &=& \iiint d^3 \vec{x}   d^3 \vec{x}_{\rm obj}   d^3 \vec{x}'_{\rm obj}   \frac{G^2 \bar{\rho}_g \rho_{\rm obj}(\vec{x}_{\rm obj}) \rho_{\rm obj} (\vec{x}'_{\rm obj})   (z-z'_{\rm obj}-\mathcal{M} c_s t)}{c_s^2   [(x-x'_{\rm obj})^2 + (y-y'_{\rm obj})^2 + (z-z'_{\rm obj}-\mathcal{M} c_s t)^2]^{3/2}} \nonumber \\ 
    &&  \times \frac{\mathcal{S}(\vec{x}, t; \vec{x}_{\rm obj})}{\{[(x-x_{\rm obj})^2 +(y-y_{\rm obj})^2] (1- \mathcal{M}^2) + (z-z_{\rm obj} - \mathcal{M} c_s t)^2\}^{1/2}}~.   
    \label{eq:FDFgeneralfinal}
\end{eqnarray}
Here, the integral over $\vec{x}_{\rm obj}$ generates the wake, which in turn exerts a drag on the mass elements at $\vec{x}_{\rm obj}'$. The deviation of EDCO friction force from the point-mass result can be parametrized as a correction contribution
\begin{equation}
I(\mathcal{M}, t) = I_{\rm pm} (\mathcal{M}, t)\big|_{r_{\rm min} = R_{\rm obj}} + I_{\rm ext} (\mathcal{M}, t)~,
\label{eq:Iextdef}
\end{equation}
where $I_{\rm pm}$ is the standard point-mass expression and $I_{\rm ext}$ encodes the finite-size correction. The correction depends on the internal mass profile, and for spherically symmetric EDCOs, it contributes only in the supersonic regimes at late times.

\begin{figure}[t]
\centering
\includegraphics[width=0.49\linewidth]{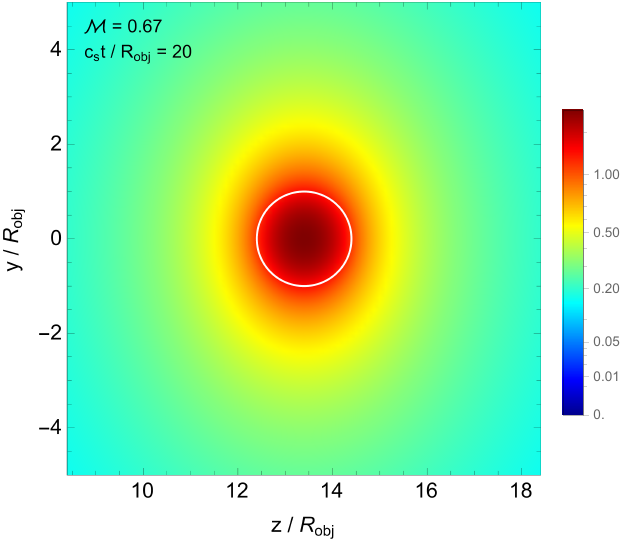}
\includegraphics[width=0.49\linewidth]{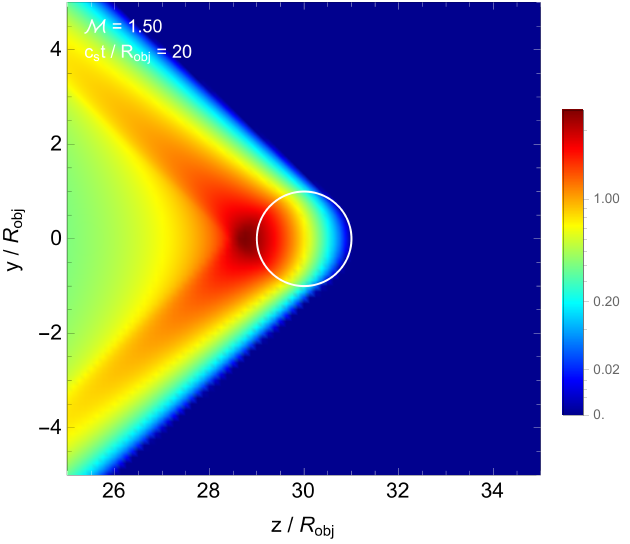}
\caption{
Zoomed-in wake profiles around a spherically symmetric EDCO with a uniform density distribution at $c_s t / R_{\rm obj} = 20$, for subsonic with $\mathcal{M} = 0.67$ [Left] and supersonic with $\mathcal{M} = 1.50$ [Right] motion. The white circle indicates the object radius $R_{\rm obj}$. In the subsonic case, the wake forms a smooth overdensity surrounding the object, while in the supersonic case, a sharp Mach cone intersects the object, leading to an enhanced and asymmetric wake structure. Since the wake in this regime depends only on the distance to the object as in Eq.~\eqref{eq:S}, the near-field profile shown here does not evolve with time.
} 
\label{fig:wakezoomedin}
\end{figure}

\subsection{Numerical evaluation of finite-size corrections}

In Figs.~\ref{fig:wakesubsonic}--\ref{fig:wakezoomedin}, we display density plots of the numerically computed wake profiles from Eq.~\eqref{eq:alphaextobj} for spherically symmetric EDCOs with a uniform density distribution considering variety of parameters as well as subsonic and supersonic cases. The objects are taken to move along the $z$-axis, and wakes extend both inside and outside EDCOs. 
In Fig.~\ref{fig:wakezoomedin}, zoomed-in wake profiles around the objects show that in the subsonic case the wake manifests as a smooth symmetric overdensity enveloping the object. By contrast, in the supersonic regime, a pronounced Mach cone cuts across the object producing an enhanced and asymmetric wake. Since the near-field perturbation depends only on the distance from the object as in Eq.~\eqref{eq:S}, its profile remains time-independent. 
Compared to the point-mass case analyzed in Ref.~\cite{Ostriker:1998fa}, the profiles are smeared over the scale of the EDCO radius $R_{\rm obj}$, reflecting finite-size structure effects.

To numerically evaluate Eq.~\eqref{eq:FDFgeneralfinal}, we discretize EDCOs into concentric coaxial rings aligned with the $z$-axis, exploiting the axial symmetry of the system. For each mass element, we then computed the wake contribution using Eq.~\eqref{eq:alphapointmass}, determined its back-reaction on EDCO, and then summed contributions of all elements. The cross-sectional size of each ring was chosen as $dl = R_{\rm obj}/40$, which ensured sufficiently stable numerical convergence of the integral. The total mass $M_{\rm obj}$ and radius $R_{\rm obj}$ were normalized across all density profiles and $I(\mathcal{M}, t)$ and $I_{\rm ext}(\mathcal{M}, t)$ were extracted from Eq.~\eqref{eq:FDFparameterize} and Eq.~\eqref{eq:Iextdef}.

\begin{figure}[t]
\centering
\includegraphics[width=0.48\linewidth]{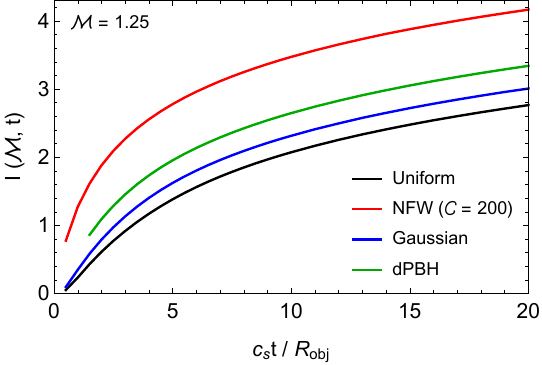}
\includegraphics[width=0.49\linewidth]{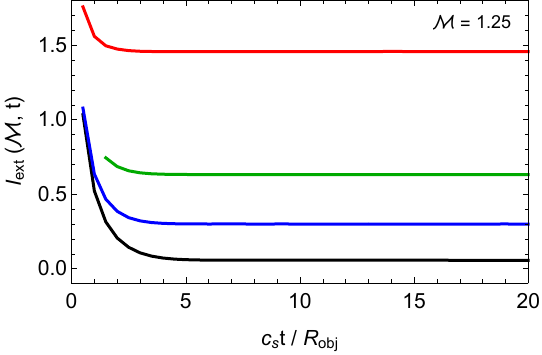} \\
\includegraphics[width=0.48\linewidth]{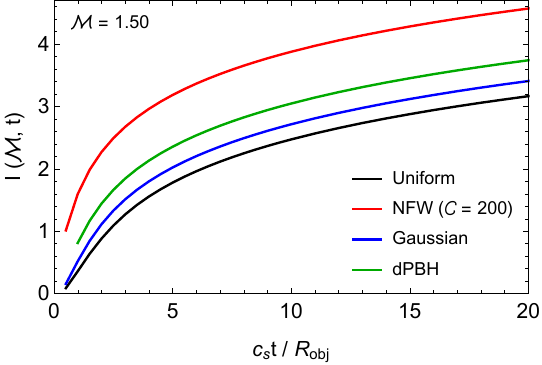}
\includegraphics[width=0.49\linewidth]{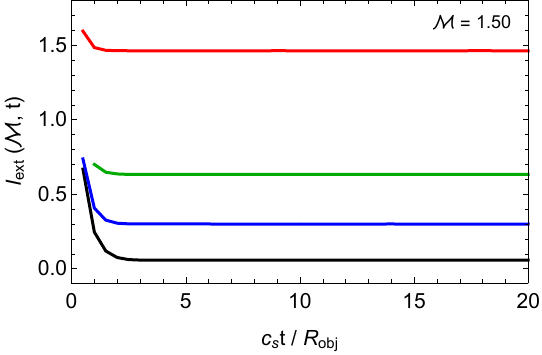}
\caption{Time evolution of $I(\mathcal{M}, t)$ [Left] and $I_{\rm ext}(\mathcal{M}, t)$ [Right] for representative EDCO density distributions with uniform (black), NFW with concentration $\mathcal{C} = 200$ (red), Gaussian (blue) and dPBH (green) profiles at Mach numbers $\mathcal{M} = 1.25$ [Top] and $\mathcal{M} = 1.5$ [Bottom]. The total factor $I(\mathcal{M}, t)$ exhibits the expected logarithmic growth with time, while the correction term $I_{\rm ext}(\mathcal{M}, t)$ converges rapidly to its asymptotic value, highlighting the finite-size dependence of the dynamical friction force.}
\label{fig:FDF}
\end{figure}

\subsection{Behavior of $I_{\rm ext}$}

In Fig.~\ref{fig:FDF} we present the time evolution of $I(\mathcal{M}, t)$ and $I_{\rm ext}(\mathcal{M}, t)$ for representative EDCOs at $\mathcal{M} = 1.25$ and $\mathcal{M} = 1.5$. The objects considered are described in Sec.~\ref{sec:selected}: Q-balls with a uniform density profile (Eq.~\eqref{eq:Qballdensity}), axion miniclusters with Navarro-Frenk-White (NFW) density profile (Eq.~\eqref{eq:NFWamc}, with concentration $\mathcal{C} = 200$), axion and dark fermion stars with a Gaussian profile (Eq.~\eqref{eq:Gaussianas}) and dPBH profile
composed of a central PBH surrounded by a DM halo (Eq.~\eqref{eq:rhoh}). The calculations were performed at intervals of $c_s t = 0.5 R_{\rm obj}$, with $t=0$ corresponding to the initial state without a wake\footnote{For dressed PBHs, the earliest points were excluded to ensure numerical stability.}.

For both subsonic and supersonic cases we find that $I_{\rm ext}(\mathcal{M}, t)$ converges to an asymptotic value once $\mathcal{M} c_s t \gtrsim R_{\rm obj}$, that is once the near-field wake has settled into a time-independent configuration. In all subsonic cases, this asymptotic value is $I_{\rm ext} = 0$. However, for supersonic cases $I_{\rm ext} > 0$ and depends on the density profile, but is numerically found to be largely independent of $\mathcal{M}$ for considered $\mathcal{M} < 3$. The obtained asymptotic results are summarized in Tab.~\ref{tab:Reff}.

The rapid convergence of $I_{\rm ext}$ reflects scale separation and can be understood as follows. The finite-size correction arises only from the near-field region $r \sim R_{\rm obj}$. At larger radii, the flow responds only to the total mass and is thus identical to the point-mass case due to gravitational shell theorem. When $c_s t \gtrsim R_{\rm obj}$, the near-field wake becomes stable and $I_{\rm ext}$ saturates to a constant value determined by the density profile.
For subsonic motion, forward-backward symmetry of the near-field cancels the finite-size contribution, resulting in vanishing $I_{\rm ext} = 0$. 
For supersonic motion, the Mach cone breaks this symmetry and produces a nonzero, profile-dependent constant $I_{\rm ext} > 0$.

Comparing density profiles, we find $I_{\rm ext}$ is larger for EDCOs with mass distributions more concentrated toward the center. Such objects, for example EDCOs with NFW profiles and large concentration $\mathcal{C}$, behave effectively as point mass objects with $r_{\rm min} \ll R_{\rm obj}$. Thus, $I_{\rm ext}$ for them can be interpreted as an enhancement of the logarithmic term in Eq.~\eqref{eq:IMachpointmass}. A full analytic analysis of the $I_{\rm ext}$ behavior is beyond our scope and left for future work. Our numerical results up to $\mathcal{M}=3$ effectively capture behavior relevant for analysis in gaseous environment of Leo T. In particular, integration of the DM Maxwell–Boltzmann velocity distribution of Eq.~\eqref{eq:MB}, with $\sigma_v=6.9$~km~s$^{-1}$ and $c_s=9$~km~s$^{-1}$, shows that around $\simeq 60\%$ of EDCOs will be supersonic  with $\mathcal{M} > 1$, and only $\mathcal{O}(0.1)\%$ of EDCOs will exceed $\mathcal{M}\geq 3$. Therefore, we adopt the asymptotic $I_{\rm ext}$ values of Tab.~\ref{tab:Reff} for our analysis and deriving dynamical friction constraints in Sec.~\ref{sec:selected}.

\begin{table}[t]
\centering
\begin{tabular}{ |l| l| l| c| c| }
    \hline
    EDCO Type & Density Profile & Density Equation & Asym.~$I_{\rm ext} (\mathcal{M} > 1)$ \\ \hline \hline
    Q-balls & Uniform & ~~Eq.~\eqref{eq:Qballdensity} & 0.057 \\ \hline
    Axion miniclusters & NFW ($\mathcal{C} = 200$) & ~~Eq.~\eqref{eq:NFWamc} & 1.460 \\ \hline 
    Axion \& dark fermion stars & Gaussian & ~~Eq.~\eqref{eq:Gaussianas} & 0.299 \\ \hline 
    Dressed PBHs & dPBH & ~~Eq.~\eqref{eq:rhoh} with PBH & 0.631 \\ \hline
\end{tabular}
\caption{Density profiles of selected representative EDCOs for Q-balls, axion miniclusters, axion and dark fermion stars and dPBHs as well as the numerically computed asymptotic corrections $I_{\rm ext}$ in the supersonic regime ($\mathcal{M} > 1$). For dPBHs, $I_{\rm ext}$ is evaluated at $M_{\rm PBH} = 10^4 M_\odot$, with negligible mass dependence in the range of interest.}
\label{tab:Reff}
\end{table}

\section{Accretion Disk Heating and Outflows}
\label{sec:disk}

Compact objects traversing the interstellar medium can accrete surrounding gas and dust, leading to the formation of accretion disks. The accretion rate depends on the enclosed mass $M_{\rm enc}$, ambient gas density $n$ and the relative velocity between the object and the medium, and is described by the Bondi–Hoyle–Lyttleton (BHL) formalism~\cite{1939PCPS...35..405H,1944MNRAS.104..273B,1952MNRAS.112..195B} 
\begin{equation}
\label{eq:bondihoyle}
\dot{M}_{\rm BHL} = \frac{4\pi G^2 M_{\rm enc}^2 n \mu m_p}{\tilde{v}^3}
\simeq 3.7\times10^{11}~{\rm g~s^{-1}}
\left(\frac{M_{\rm enc}}{M_\odot}\right)^2
\left(\frac{n}{1\textrm{ cm}^{-3}}\right)
\left(\frac{\tilde{v}}{10\textrm{ km/s}}\right)^{-3},
\end{equation}
where $\tilde{v}=\sqrt{v^2+c_s^2}$ incorporates the sound speed $c_s\simeq 9$~km~s$^{-1}$, $m_p \simeq 938.3$~MeV is the proton mass, and $\mu\simeq 1$ is the typical mean molecular weight for atomic hydrogen-dominated clouds. The Bondi radius  inside which material is gravitationally captured is given by
\begin{equation}
\label{eq:rbondi}
R_{\rm Bondi} = \frac{2GM_{\rm enc}}{\tilde{v}^2}
\simeq 8.6\times10^{-5}~{\rm pc}
\left(\frac{M_{\rm enc}}{M_\odot}\right)
\left(\frac{\tilde{v}}{10\textrm{ km/s}}\right)^{-2}.
\end{equation}
For dPBHs, which consist of a central PBH surrounded by a DM halo, $M_{\rm enc}$ represents the total enclosed mass within $R_{\rm Bondi}$ that exceeds the bare black hole mass $M$. For other EDCOs that are typically smaller than $R_{\rm Bondi}$, one has $M_{\rm enc}=M$.

The gravitational potential energy of accreted matter powers radiation from the accretion disk with luminosity
\begin{equation}
L = \epsilon \dot{M}c^2,
\end{equation}
where $\epsilon$ is the radiative efficiency. The efficiency depends primarily on the innermost stable circular orbit (ISCO); the smaller the ISCO radius, the larger the fraction of gravitational energy released. For non-rotating Schwarzschild black holes\footnote{This is modified for rotating (Kerr) black holes, which we do not consider.}, $R_{\rm ISCO}=3R_s$ and the maximum efficiency is $\epsilon \simeq 0.057$~\cite{2008bhad.book.....K}. However, the efficiency can be significantly lower depending on the accretion rate and the nature of the accretion flow. In addition to dPBHs, as described in Sec.~\ref{sec:selected}, we consider variety of EDCOs that extend beyond black holes with effective inner radii $r_{\rm min}>3R_s$, which we take to be their physical radii. While most semi-analytic accretion models were derived for black holes, their explicit scalings with $r_{\rm min}$ enable us to generalize the results to EDCOs with different $r_{\rm min}$. For transparent EDCOs, accretion flows within objects themselves can in principle also occur, however, analysis of such possible processes is beyond the scope of this work. Here, we consider negligible photon emission or outflows from the interior of EDCOs.

The photon emission efficiency of an accretion flow increases as the accretion rate approaches the Eddington limit, where radiation pressure balances gravitational attraction considering spherical symmetry. The corresponding Eddington accretion rate is
\begin{equation}
\dot{M}_{\rm Edd} \simeq 1.3\times10^{18}~{\rm g~s^{-1}} \left(\frac{M}{M_\odot}\right),
\end{equation}
assuming a reference radiative efficiency of $\epsilon = 0.1$. Super-Eddington flows, such as the slim disk solutions discussed in Sec.~\ref{ssec:slim}, can exceed this rate through the coexistence of inflows and outflows.
 It is useful to compare the BHL accretion rate to the Eddington rate by defining the dimensionless accretion parameter 
\begin{equation}
\label{eq:dimmdot}
\dot{m} = \frac{\dot{M}_{\rm BHL}}{\dot{M}_{\rm Edd}}
\simeq 2.64\times10^{-7}\left(\frac{M_{\rm enc}}{M_\odot}\right)^2\left(\frac{M}{M_\odot}\right)^{-1}\left(\frac{n}{1~{\rm cm^{-3}}}\right)\left(\frac{\tilde{v}}{10~{\rm km~s^{-1}}}\right)^{-3}~,
\end{equation}
where the Bondi rate depends on the enclosed mass at the Bondi radius $M_{\rm enc}$, while the Eddington rate is set by the central object mass $M$ that governs the accretion disk dynamics. In the case of dPBH, $M_{\rm enc}$ is distinct from $M$. For other EDCOs that we consider, they are sufficiently compact so that effectively $M_{\rm enc} \simeq M$.

Following the classification of Ref.~\cite{Yuan:2014gma}, we characterize the accretion flow regimes according to $\dot{m}$:
\begin{itemize} [itemsep=0.1em]
    \item ADAF (advection-dominated accretion flow): $\dot{m}<7\times10^{-3}$, forming a radiatively inefficient, thick disk~\cite{Narayan:1994xi,Mahadevan:1996jf}.
    \item Thin disk: $7\times10^{-3}\lesssim\dot{m}\lesssim 1$, forming the standard geometrically thin, radiatively efficient disk~\cite{1972A&A....21....1P,Shakura:1972te}.
    \item Slim disk: $\dot{m}\gtrsim 1$, representing super-Eddington flows with significant outflows~\cite{Abramowicz:1988sp,Abramowicz:2011xu}.
\end{itemize}
We adopt semi-analytic models for these three regimes in the following discussion.

We next consider an approximate criterion for accretion disk formation—namely, whether angular momentum accreted from the interstellar medium is sufficient to sustain a disk around an EDCO. Following Ref.~\cite{Agol:2001hb}  and assuming a Kolmogorov spectrum of density fluctuations, the inflowing gas acquires a fractional density perturbation $\delta \rho/\rho \sim (R_{\rm Bondi}/10^{18}{\rm cm})$ at the Bondi radius, which corresponds to a characteristic angular momentum. Equating this to the Keplerian angular momentum at the outer edge of a thin disk $r_{\rm disk}$ yields the condition that $r_{\rm disk}$ must exceed the innermost stable circular orbit (ISCO) or physical radius of the object $r_{\rm min}$. This gives the approximate criterion 
\begin{equation}
\label{eq:thindiskcond}
M > 2.5\times10^{-13}~M_\odot \left(\frac{\tilde{v}}{10~{\rm km~s^{-1}}}\right)^{5} \left(\frac{r_{\rm min}}{3R_s}\right)^{3/2},
\end{equation}
where $r_{\rm min}$ is the ISCO radius for black holes and the physical radius for non–black hole EDCOs. Physically, this condition considers that turbulence-driven angular momentum must be sufficiently large in order to maintain orbits beyond $r_{\rm min}$ and allowing a disk to form.

Although these results are derived for thin-disk formation, similar considerations can be expected to apply to thick and slim disks as well. The constraint on the mass-radius relation of Eq.~\eqref{eq:thindiskcond} is well satisfied for intermediate mass PBHs considered in Sec.~\ref{sec:dpbh} and also for significant parameter space of dark fermion stars and Q-balls, as can be seen from Fig.~\ref{fig:compactranges}. On the other hand, axion miniclusters and axion stars fail to satisfy the criterion of Eq.~\eqref{eq:thindiskcond} in parts of their mass range parameter space, signifying that accretion disks are unlikely to form in those cases. Moreover, accretion-disk processes become negligible for extended objects with $R \gg 100R_s$ long before disk formation ceases to be possible. In this regime, dynamical friction bounds, as shown in Fig.~\ref{fig:fallowedvary}, dominate and provide more robust constraints than those based on accretion-disk physics.

\subsection{Thick disks}
\label{ssec:thick}
 
To model thick ADAF disks, we adopt the semi-analytic framework of Ref.~\cite{Mahadevan:1996jf} as implemented in Refs.~\cite{Lu:2020bmd,Takhistov:2021aqx}. Here, we summarize the key ingredients and formalism (see Refs.~\cite{Mahadevan:1996jf,Takhistov:2021aqx} for additional details). Within this framework, the ADAF regime is further divided into sub-classes according to the dimensionless accretion rate: (electron) eADAF for $\dot{m} < 10^{-5}$, ADAF for $10^{-5}<\dot{m}<10^{-3}$ and (luminous hot accretion flow) LHAF for $10^{-3}<\dot{m}<7\times10^{-3}$. These regimes share similar qualitative disk structure but represent different breakpoints in radiative photon efficiency. Throughout, we take the viscosity parameter $\alpha=0.1$ and the gas to pressure ratio $\beta=10/11$, along with fiducial parameters used in Refs.~\cite{Lu:2020bmd,Takhistov:2021aqx}. 

ADAF disks are turbulent, geometrically thick and optically thin, producing a complex multi-component emission spectrum that includes synchrotron, inverse Compton (IC), and bremsstrahlung radiation. Synchrotron emission dominates at low frequencies with $L_\nu \propto \nu^{5/2}$ up to a peak at 
\begin{equation}
    h\nu_p = 18.2 \textrm{ eV}~ \theta_e^2 \left(\frac{r_{\rm min}}{3 R_s}\right)^{-5/4}\left(\frac{M}{M_\odot}\right)^{-1/2}\left(\frac{\dot{m}}{10^{-4}}\right)^{3/4}~,
\end{equation}
where $h$ is Planck's constant.
The corresponding luminosity is
\begin{equation}
\label{eq:Lnup}
    L_{\nu_p} = 2.66\times10^{32}~  \textrm{erg~eV$^{-1}$~s$^{-1}$}\left(\frac{r_{\rm min}}{3 R_s}\right)^{-1/2}\left(\frac{M}{M_\odot}\right)\left(\frac{\dot{m}}{10^{-4}}\right)^{3/2}~.
\end{equation}
Here, $\theta_e = k_B T_e/m_e c^2$ is the dimensionless electron temperature, where $k_B$ is the Boltzmann constant and $c$ the speed of light that we restore for clarity, for electron temperature $T_e$, electron mass $m_e = 0.511$~MeV, determined self-consistently by equating electron heating and cooling within the disk plasma\footnote{This is distinct from the interstellar medium gas heating.}.

IC scattering of synchrotron photons produces a power-law tail $L_{\nu,\rm IC} \propto L_{\nu_p} (\nu/\nu_p)^{-\alpha_c}$ for intermediate frequencies, $h\nu_p<h\nu<3 k_B T_e$. The spectral index is 
\begin{equation}
\alpha_c = \dfrac{\ln\tau_{\rm es}}{\ln A}~,
\end{equation} 
where the electron scattering optical depth is
\begin{equation}
    \tau_{\rm es} = 7.2\times10^{-3}\left(\frac{r_{\rm min}}{3 R_s}\right)^{-1/2} \left(\frac{\dot{m}}{10^{-4}}\right)~,
\end{equation}
and $A = 1 + 4\theta_e + 16\theta_e^2$.

Bremsstrahlung emission dominates at high energies near $h \nu\sim k_B T_e$, with an exponential cutoff 
\begin{equation}
    L_{\nu,\rm brems}(\nu) \simeq 3.73\times10^{23} ~  \textrm{erg~eV$^{-1}$~s$^{-1}$}~ \theta_e^{-1} F(\theta_e)e^{-h\nu/k_B T_e} \ln\left(\frac{10^3 R_s}{r_{\rm min}}\right)\left(\frac{M} {M_\odot}\right)\left(\frac{\dot{m}}{10^{-4}}\right)^2
    \label{eq:lbrems}
\end{equation}
with $F(\theta_e)\sim\mathcal{O}(1)$. These scalings are expected to hold for EDCOs with $R \lesssim 10^3 R_s$. For more extended objects, the approximations, such as the logarithm in Eq.~\eqref{eq:lbrems}, break down. However, these limitations do not significantly restrict our analysis, since for $R\gtrsim100 R_s$ the heating from accretion disk photons is already subdominant to that from dynamical friction.

In addition to radiative photon emission, ADAF simulations indicate the presence of significant outflows~\cite{Yuan:2014gma}. The outflow rate can be parametrized as~\cite{Blandford:1998qn,Totani:2006zx}
\begin{equation}
\dot{M}_{\rm wind}(r) = s \dot{M}_{\rm BHL}\left(\frac{r}{r_{\rm out}}\right)^s~,
\end{equation}
with slope $s$, from the inner radius $r_{\rm in}$ to the outer radius $r_{\rm out}$. For EDCOs, we have verified that the ``high outflow'' prescription of Ref.~\cite{Takhistov:2021aqx} typically yields $r_{\rm out}$ smaller than the object radius. Hence, for our analysis we conservatively adopt the ``low outflow'' model with $s=0.5$, $r_{\rm in}=\max(10 R_s, r_{\rm min})$, $r_{\rm out}=R_{\rm Bondi}$. The kinetic energy of outflowing protons at infinity is modeled as a fraction $f_k=0.1$ of the Keplerian velocity at the launch radius, producing an energy spectrum 
\begin{align}
\label{eq:fethick}
    f(E) =&~ 1.57\times10^{25} ~  \textrm{erg~eV$^{-2}$~s$^{-1}$}~ \left(\frac{E}{\textrm{MeV}}\right)^{-3/2}\left(\frac{M}{M_\odot}\right)^2\left(\frac{n}{1\textrm{ cm}^{-3}}\right)\left(\frac{\tilde{v}}{10\textrm{ km~s$^{-1}$}}\right)^{-3} \notag\\ & \times\left(\frac{r_{\rm out}}{R_s}\right)^{-0.5}\left(\frac{f_k}{0.1}\right)~,  
\end{align}
valid over $2.35\times10^{-2}\textrm{ eV}~(r_{\rm out}/10^8R_s)^{-1}<E<235\textrm{ keV}~(r_{\rm in}/10R_s)^{-1}$.
 
Thick ADAF disks are radiatively inefficient and most of the gravitational binding energy of the accreted matter is stored as thermal energy in the hot ion plasma and advected inward rather than radiated away. As a result, ADAFs produce relatively dim emission compared to thin or slim disks at the same accretion rate, but can 
launch energetic outflows. This inefficiency is relevant for our constraints, since it limits the photon heating contribution from EDCOs, making dynamical friction effects comparatively more important for large radii when $R\gtrsim 100R_s$.

\subsection{Thin disks}
\label{ssec:thin}

For higher sub-Eddington accretion rates, $7\times10^{-3}\lesssim \dot{m} \lesssim 1$, the flow settles into a geometrically thin optically thick disk. In this regime, the disk radiates approximately as a superposition of blackbodies at different radii. The temperature reaches its maximum near the inner edge $r_{\rm min}$ and decreases outward, producing a multi-temperature blackbody spectrum.

Thin accretion disk admits an analytic description. The temperature profile for $r_{\rm min}< r < r_{\rm out}$ is
\begin{equation}
T(r) = T_i \left(\dfrac{r_{\rm min}}{r}\right)^{3/4}\left[1-\left(\dfrac{r_{\rm min}}{r}\right)^{1/2}\right]^{1/4}~,
\end{equation}
with a maximum value $T_{\rm max} \simeq 0.488 T_i$ attained at $r=(49/36)~ r_{\rm min}$. The corresponding blackbody spectrum peaks at photon energies $h \nu_p \sim 2.8 k_B T_{\rm max}\sim k_B T_i$.

The spectral range is determined by two characteristic temperatures~\cite{Pringle:1981ds}. The first is the characteristic temperature of the inner disk 
\begin{equation}
k_B T_i \simeq 53.3~{\rm eV}~\left(\dfrac{n}{1~{\rm cm}^{-3}}\right)^{1/4}\left(\dfrac{\tilde{v}}{10~{\rm kms}^{-1}}\right)^{-3/4}\left(\frac{r_{\rm min}}{3 R_s}\right)^{-3/4}\left(\dfrac{M_{\rm enc}}{M}\right)^{1/2}~.
\end{equation}
The second is the outer-disk temperature at $r \sim R_{\rm Bondi}$ 
\begin{equation}
k_B T_o \simeq 6\times10^{-5}{\rm eV}~\left(\frac{n}{1{\rm cm}^{-3}}\right)^{1/4}\left(\dfrac{M_{\rm enc}}{M_\odot}\right)^{-1/2}\left(\dfrac{\tilde{v}}{10~{\rm km~s}^{-1}}\right)^{7/4}~.
\end{equation}

\begin{figure}[t]
\centering
\includegraphics[width=0.7\linewidth]{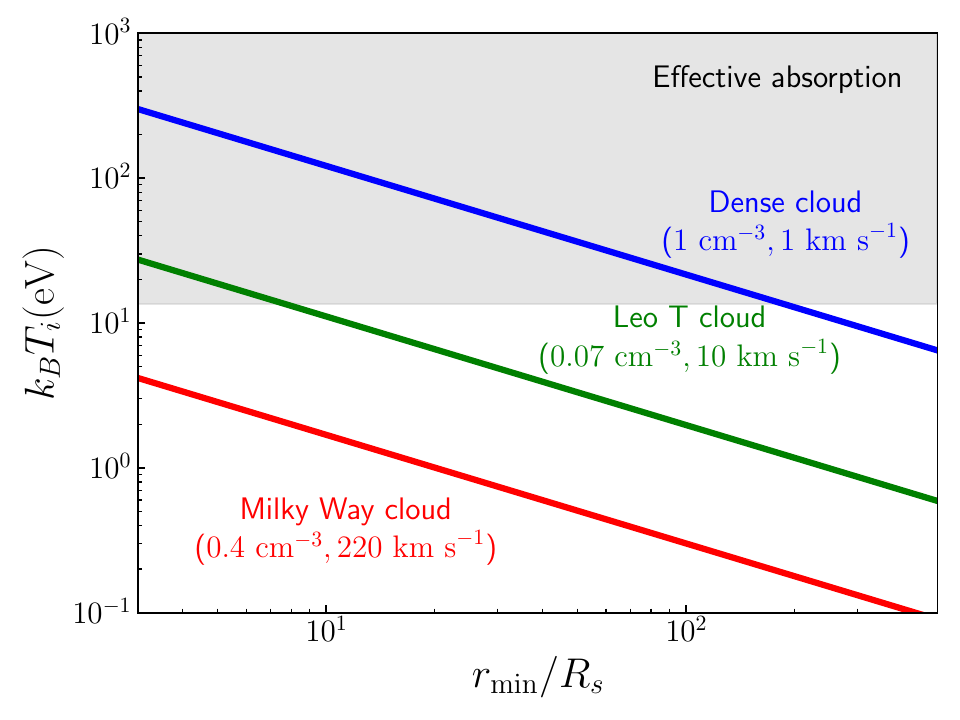}
\caption{Characteristic inner temperature $T_i$ of thin accretion disk as a function of EDCO $r_{\rm min}$. Parameters representative of Milky Way gas clouds (red,  $n=0.4~\textrm{cm}^{-3}$ and $\tilde{v}=220~\textrm{km~s}^{-1}$), Leo T (green,  $n=0.07~\textrm{cm}^{-3}$ and $\tilde{v}=10~\textrm{km~s}^{-1}$), and a dense gas cloud ($n=1~\textrm{cm}^{-3}$ and $\tilde{v}=1~\textrm{km~s}^{-1}$). As the object radius increases, the peak photon energy $\sim k_B T_i$ falls below the $13.6~\textrm{eV}$ ionization threshold of atomic hydrogen (gray shaded region), reducing absorption and the efficiency of gas heating. The exponentially suppressed high energy tail of the spectrum beyond the peak can still contribute appreciably to interstellar medium heating. We set $M_{\rm enc}=M$ throughout.}
\label{fig:peakfrequency}
\end{figure}

The resulting spectrum consists of three regions: a Rayleigh-Jeans tail with $L_\nu \propto \nu^2$ for $h\nu<k_B T_o$, an intermediate slope with $L_\nu \propto \nu^{1/3}$ for $k_B T_o<h\nu<k_B T_i$, and an exponential cutoff for $h\nu>k_B T_i$, with $L_\nu \propto \nu^2 e^{1-h\nu/k_B T_i}$. This can be captured by the piecewise approximation~\cite{Pringle:1981ds,Takhistov:2021aqx} as
\begin{equation} \label{eq:thinspectrum} L(\nu) \simeq \begin{cases} c_\alpha \left(\dfrac{T_{\rm i}}{T_{o}}\right)^{5/3} \left(\dfrac{h\nu}{k_B T_{\rm i}}\right)^2, & \text{for $h\nu < k_BT_{o}$}~,\\ c_\alpha \left(\dfrac{h\nu}{k_BT_{i}}\right)^{1/3}, & \text{for $k_B T_o <h\nu < k_BT_{i}$}~,\\ c_\alpha \left(\dfrac{h\nu}{k_BT_{\rm i}}\right)^2 e^{1-h\nu/k_BT_{i}}, & \text{for $k_BT_{i} < h\nu$}~, \end{cases} \end{equation}
with  normalization 
\begin{equation} c_\alpha = 3.09 \times 10^{33} \textrm{ erg}\textrm{ eV}^{-1} ~{\rm s}^{-1}~\dot{m}^{2/3} \left(\frac{M}{M_\odot}\right)^{4/3} \left(\frac{n}{1\textrm{ cm}^{-3}}\right)^{1/12}\left(\frac{\Tilde{v}}{10\textrm{ km~s$^{-1}$}}\right)^{-1/4}~. \end{equation}

In Fig.~\ref{fig:peakfrequency} we illustrate the characteristic inner disk temperature for representative environments: Milky Way gas clouds (red) as considered in Refs.~\cite{Wadekar:2019xnf,Takhistov:2021aqx}, Leo T dwarf galaxy (green), and a dense cloud with low relative velocity (blue). 
For EDCOs with $r_{\rm min}\lesssim 100 R_s$, a significant fraction of the emission lies above the Lyman limit  $E_i = 13.6$ eV, generating ionizing UV photons that can be efficiently absorbed by surrounding atomic hydrogen. Unlike ADAFs and slim disks, thin disks are radiatively efficient and are not expected to drive strong winds or outflows. 
For Leo T, while the inner temperature falls below the ionization threshold and efficient gas absorption (gray shaded), the exponential tail of the spectrum still provides efficient heating for EDCOs with inner radii up to $\sim 100 R_s$ (see Fig.~\ref{fig:fallowedvary}, left panel).

\subsection{Slim disks}
\label{ssec:slim}

EDCOs embedded in the Leo~T gas cloud can generally reach super-Eddington accretion rates $\dot{m}>1$ only for masses $M \gtrsim 10^{6} M_\odot$, as seen from Eq.~\eqref{eq:dimmdot}. This regime was therefore not considered in Refs.~\cite{Lu:2020bmd,Takhistov:2021aqx} analyzing PBHs, and is analogously not relevant for other EDCOs that we consider in this work besides dPBHs. For dPBHs, discussed in Sec.~\ref{sec:dpbh}, the surrounding DM halo substantially increases the enclosed mass within the Bondi radius, thereby enhancing the dimensionless accretion rate $\dot{m}$. Thus, unlike other EDCOs, dPBHs with central black hole masses as small as $M\gtrsim 10^{2} M_\odot$ can enter the super-Eddington regime. We model these cases using the slim disk solution, which incorporates both photon emission and energetic outflows.

For the photon component, we adopt the semi-analytic prescription of Ref.~\cite{Takeo:2019uef}, which treats the slim disk as a modified thin disk. In this regime, the inner edge of the disk moves inward, but the overall radiative efficiency decreases due to photon trapping. The luminosity saturates at $L \to 2 L_{\rm Edd}$ in the $\dot{m}\to\infty$ limit. This approach neglects optically thin emission from the hot inner region, which is both subdominant and inefficiently absorbed by the interstellar gas. To ensure a smooth transition between the thin disk and slim disk regimes, we adapt the equations of Ref.~\cite{Takeo:2019uef} to match our definition of the Eddington accretion rate, $\dot{M}_{\rm Edd}=0.1L_{\rm Edd}/c^2$, with continuity enforced at $\dot{m}=1$.

The resulting slim disk spectrum resembles that of a thin disk, but with an additional suppression inside the photon-trapping radius where radiation is advected with the inflow. This introduces a cutoff frequency corresponding to the effective temperature at the trapping radius 
\begin{equation}
    \nu_{\rm tr} = 3.34\times10^{16}\textrm{ Hz}~ \left(\frac{M_{\rm PBH}}{10~ M_\odot}\right)^{-1/4}\left(\frac{\dot{m}}{10^3}\right)^{-1/2}~.
\end{equation}
The slim disk spectrum can then be described as a modified thin disk spectrum of Eq.~\eqref{eq:thinspectrum},  with an additional regime for $\nu_{\rm tr}<\nu<T_i$ 
\begin{equation}
    L_\nu(\nu) = 1.70\times10^{34} \textrm{ erg~}\textrm{eV}^{-1}~{\rm s}^{-1}  ~\left(\frac{\nu}{\nu_{\rm tr}}\right)^{-1}\dot{m}^{1/2} \left(\frac{M_{\rm PBH}}{M_\odot}\right)^{5/4}~.
\end{equation} 
Thus, the luminosity peak shifts from $\nu\sim T_i$ to $\nu\sim \nu_{\rm tr}$, and the high-frequency tail above $\nu>T_i$ is reduced to match the value at $L_\nu(T_i)$.

For the outflow component, we use the semi-analytic model of Refs.~\cite{Yang:2022cpm,Yang:2024nte}, which describes energetic winds launched from the slim disk. Similar to the ADAF case, the outflow rate can be approximated by a power law model with $s=0.83$\footnote{We included an additional 0.5 prefactor compared to the ADAF case, accounting for half of the accretion being expelled as outflow, so that the energy of the winds is approximately $1\%$ of the inflowing mass-energy matching Refs.~\cite{Yang:2022cpm,Yang:2024nte}.}
\begin{equation}
    \dot{M}_{\rm wind}(r) = 0.42 \dot{M}_{\rm BHL} \left(\frac{r}{10^2 R_s}\right)^{0.83}~,
\end{equation}
with $r_{\rm in}=\max(10R_s,r_{\rm min})$ and $r_{\rm out}=10^2 R_s$. In this model, roughly half of the inflow at the Bondi radius is expelled in the wind. The wind velocities are much higher than in the ADAF case, with poloidal and toroidal components
\begin{align}
\label{eq:windvel}
    & v_p(r) = 0.15 c~, \\
    & v_\phi (r) = 0.79 v_k(r)~,
\end{align}
where $v_k=\sqrt{GM/r}$ is the Keplerian velocity. For comparison, ADAF winds typically have $v\sim 0.1 v_k$. The kinetic power of slim disk winds is dominated by the poloidal component and can be estimated as $\dot{E}_{\rm wind}\sim 10^{-2} \dot{M}_{\rm BHL}c^2$.

The wind emission particle energy at a given radius is
\begin{equation}
E(r)=\dfrac{\mu m_p}{2} \sqrt{(0.15c)^2+(0.79v_k)^2}~.
\end{equation}
This yields a power law energy distribution 
  parameterized as  
\begin{align}
\label{eq:feslim}
    f(E) =&~ 2.93\times10^{28}~\textrm{erg}~\textrm{eV}^{-2}\textrm{ s}^{-1}  ~\left(\frac{E}{\textrm{MeV}}-10.6\right)^{-1.83}\left(\frac{M_{\rm enc}}{M_\odot}\right)^2 \notag\\
    &\times \left(\frac{n}{1\textrm{ cm}^{-3}}\right)\left(\frac{\tilde{v}}{10\textrm{ km/s}}\right)^{-3}\left(\frac{r_{\rm out}}{10^2 R_s}\right)^{-0.83}~, 
\end{align}
valid for $10.6~{\rm MeV}~\lesssim E < 10.6+14.7(r_{\rm in}/10R_s)^{-1}~{\rm MeV}$. Although slim disk winds are energetic, we find that photon emission remains the dominant source of gas heating in this regime for dPBHs.

\subsection{Optical depth, stopping power and duty cycles}
\label{ssec:addfactors}

To convert the photon emission and outflows discussed in the previous subsections into effective heating rates, we account for three additional factors: photon absorption, proton stopping power, and accretion duty cycles. We follow the methodology of Refs.~\cite{Lu:2020bmd,Takhistov:2021aqx}.

The absorption of photons from the accretion disk is set by the optical depth of the surrounding gas 
\begin{equation}
\label{eq:tau}
    \tau = n\sigma(E) r_{\rm sys}~,
\end{equation}  
where $\sigma(E)$ is the photo-ionization cross-section and for Leo T we adopt $n\simeq 0.07~{\rm cm}^{-3}$ as the hydrogen number density and $r_{\rm sys}\simeq 350~{\rm pc}$ as the characteristic size of the gas cloud. Photons with energies below $E\lesssim 10~{\rm eV}$ are non-ionizing and essentially transparent in Leo T. We therefore include only the ionizing component of the emission.  For $10~{\rm eV}<E<30~{\rm eV}$, the cross-section is described analytically as~\cite{Bethe1957,1990A&A...237..267B}
\begin{equation}
    \sigma(E) = \sigma_0 y^{-3/2} \left(1+y^{1/2}\right)^{-4}~,
\end{equation}
with $y=2E/E_i$ and $\sigma_0 = 606\textrm{ Mb} = 6.06\times10^{-16}\textrm{ cm}^2$. For higher energies, we consider the tabulated cross-section from Ref.~\cite{Olive_2014}.  The corresponding heating rate from photons is then
\begin{equation}
\label{eq:photonheating}
    \mathcal{H}_{\rm phot}(M,v) = \int_{E_{\rm min}}^{E_{\rm max}} L_\nu(M,v)f_h\left(1-e^{-\tau}\right) d\nu~,
\end{equation}
where $f_h\simeq 1/3$ is the fraction of absorbed photon energy converted into heat rather than re-radiation~\cite{1985ApJ...298..268S,Ricotti:2001zf,Furlanetto:2009uf}.

For outflowing protons, stopping power $S(E)$ quantifies the rate of energy deposition per unit column density. We employ here the tabulated values of Ref.~\cite{2019PhRvA..99d2701B}. Unlike the exponential absorption for photons, the energy deposited by a proton of initial energy $E$ is 
\begin{equation}
\Delta E \simeq \min\left(n S(E) r_{\rm sys}, E\right),
\end{equation}
so that high energy protons that traverse the system deposit only a portion of their energy. The heating rate from winds is therefore obtained by integrating over the particle spectrum given by Eq.~\eqref{eq:fethick} and Eq.~\eqref{eq:feslim}, as 
\begin{equation}
\label{eq:windheating}
    \mathcal{H}_{\rm wind}(M,v) = \int_{E(r_{\rm in})}^{E(r_{\rm out})} f_h \Delta E f(E) dE~,
\end{equation}
where $f(E)$ is the energy distribution of the outflowing protons and the limits depend on the object mass $M$ and velocity $v$.

Finally, both photon and wind heating can be suppressed by the time-dependent duty cycle of accretion, since accretion does not persist continuously over a Hubble time. We estimate the duty cycle by comparing the crossing time of the EDCO through Leo~T $t_{\rm dyn}=r_{\rm sys}/v$ with the free-fall time at the Bondi radius $t_{\rm ff} = GM/\sqrt{2}v^3$, 
\begin{equation}
    \mathcal{D} = \min\left[1,\frac{t_{\rm dyn}}{t_{\rm ff}}\right]= \min\left[1,\frac{\sqrt{2}r_{\rm sys}v^2}{GM_{\rm enc}}\right]~,
\end{equation}
including it as a multiplicative factor in Eq.~\eqref{eq:HM}.

\section{Constraints on Extended  Dark Matter Compact  Objects}
\label{sec:selected}

\begin{figure}
\centering
\includegraphics[width=0.6\linewidth]{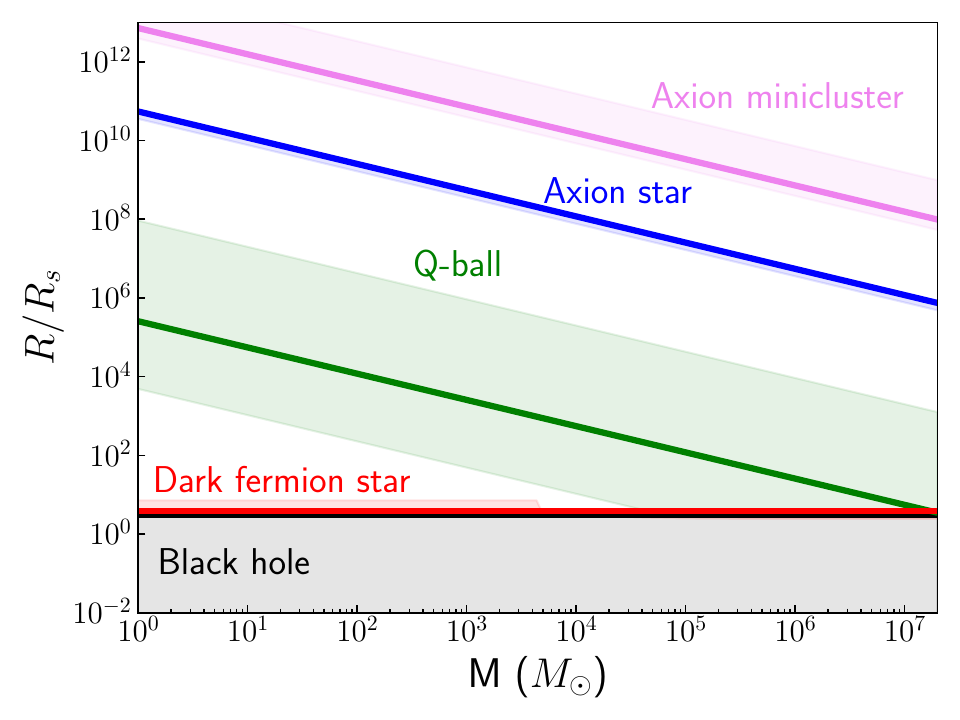}
\caption{Mass–radius relations of EDCOs expressed in units of the Schwarzschild radius $R_s$. Shown are dark fermion stars (red), Q-balls (green), axion stars (blue), and axion miniclusters (violet). For comparison, the ISCO radius $R=3R_s$ of a Schwarzschild black hole as a benchmark for accretion disk formation and maximum radiative efficiency is indicated in black. Solid lines denote the reference mass–radius relations adopted in our analysis, while shaded bands indicate the considered parameter ranges discussed in Sec.~\ref{sec:selected}.} 
\label{fig:compactranges}
\end{figure}

We analyze the impact of different classes of EDCOs on interstellar gas heating: dPBHs, axion miniclusters, axion stars, Q-balls  and dark fermion stars. These objects vary in their mass profiles and characteristic radii, as summarized in Tab.~\ref{tab:Reff}. For each scenario, we adopt a representative model together with reference parameter ranges, with the corresponding mass–radius relations shown in Fig.~\ref{fig:compactranges}. In specifying these ranges, we account for microphysical and model constraints on the underlying dark-sector fields. The sharply contrasting properties of different EDCOs lead to qualitatively distinct constraints, as displayed in Figs.~\ref{fig:falloweddhcompare}, Fig.~\ref{fig:selectedbounds}, and~Fig.~\ref{fig:fallowedcompareleotco}, which also illustrate the different heating mechanisms that dominate in each case. Since these EDCOs lack event horizons and we consider weak or negligible couplings to baryons, accreted gas can stream through the interior beyond their $r_{\rm min}$.

While our primary focus is on gas heating effects, we note that EDCOs can also be probed and constrained by a range of complementary approaches, whose detailed evaluation is beyond the scope of this work. For dPBHs, we compare our gas heating limits in Fig.~\ref{fig:falloweddhcompare} to those for bare PBHs, noting that the presence of surrounding DM halos could modify some of these bounds 
such as those that are dynamical and related to accretion. For other EDCOs, their extended radii reduce compactness relative to black holes, which tends to suppress accretion efficiency,
and could also modify constraints that assume point-like objects. Among complementary probes, bounds originating from X-rays and cosmic microwave background (CMB) can be sensitive to the physical sizes of the objects, while stellar-mass EDCOs with mass-radius relations comparable to PBHs might also be probed through gravitational wave signals from binary mergers.

\subsection{Dressed primordial black holes}
\label{sec:dpbh}

PBHs formed in the early Universe can accrete surrounding DM during the matter-dominated era, leading to the formation of an extended halo ``dress'' and producing dPBHs. Since stellar-mass PBHs are constrained over much of parameter space allowing them to constitute only a subdominant fraction of DM (e.g.~\cite{Sasaki:2018dmp,Carr:2020gox}), a large ambient DM reservoir remains available for accretion. For cold and collisionless DM, such accretion can efficiently proceed after matter–radiation equality, making the formation of dPBHs a generic outcome in a wide range of scenarios. Recently, it has been demonstrated that dPBHs can serve as unique probes of scenarios with mixed PBH and particle DM~\cite{GilChoi:2023ahp}.

The mass of the dPBH DM halo grows linearly with the scale factor during the matter-dominated era, starting from matter–radiation equality at redshifts $z\sim 3000$ and continuing until the time when nonlinear structure formation can disrupt it~\cite{Mack:2006gz,Berezinsky:2013fxa,Adamek:2019gns,Boudaud:2021irr,Oguri:2022fir}. Taking the lower cut-off redshift to be $z_c=30$, the resulting halo mass is
\begin{equation}
\label{eq:Mh}
    M_h = 97 M_{\rm PBH} \left(\frac{1+z_c}{31}\right)^{-1}~,
\end{equation}
with density profile~\cite{1985ApJS...58...39B,Berezinsky:2013fxa,Boudaud:2021irr} 
\begin{equation}
\label{eq:rhoh}
    \rho_h(r) = 0.259 M_\odot \textrm{ pc}^{-3} \left(\frac{R_h}{r}\right)^{9/4}\left(\frac{1+z_c}{31}\right)^3 ~,
\end{equation}
for $r <R_h$. 
Here, the halo radius is
\begin{equation}
\label{eq:rh}
    R_h = 0.613\textrm{ pc} \left(\frac{M_h}{M_\odot}\right)^{1/3} \left(\frac{1+z_c}{31}\right)^{-1}~.
\end{equation}

\begin{figure}[t]
\centering
\includegraphics[width=0.49\linewidth]{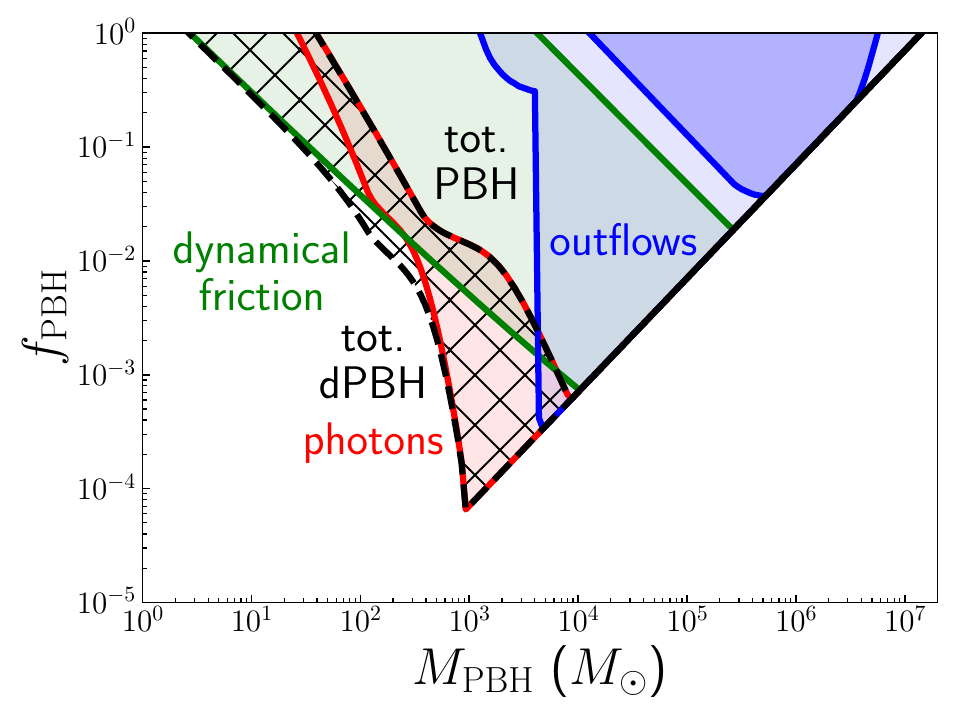}
\includegraphics[width=0.49\linewidth]{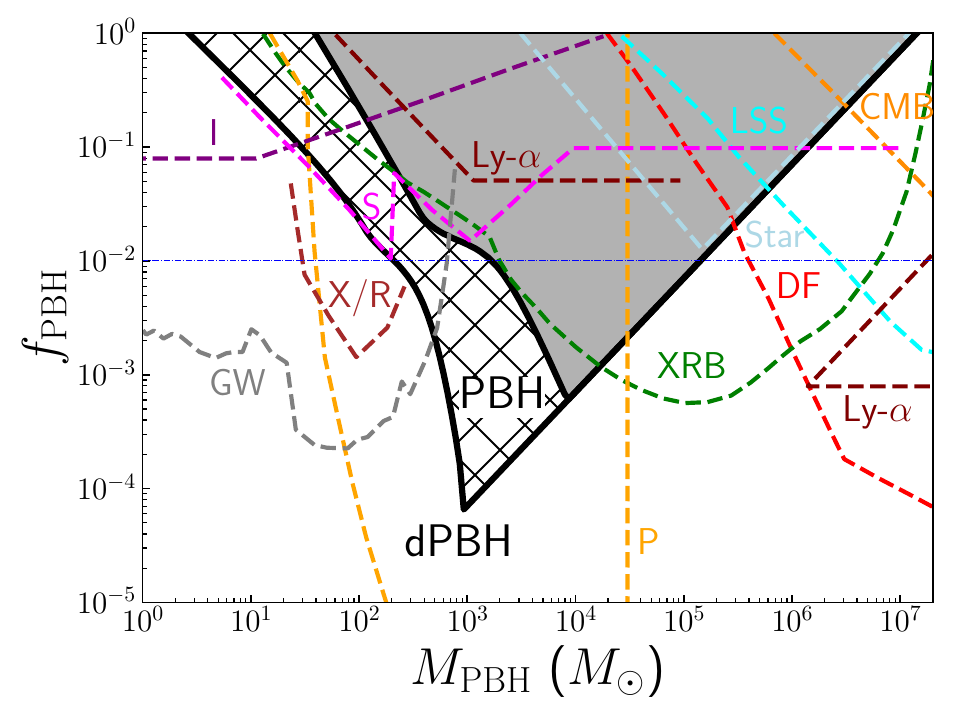}
\caption{ 
[Left] 
Interstellar gas heating constraints derived in this work for dPBHs showing the individual contributions from photon emission (red), dynamical friction (green), and outflows (blue), together with the combined bound (black dashed). The shaded regions for each component span between the stronger dPBH bounds and the weaker constraints for bare PBHs without DM halos (hatched regions between the bare PBH and dPBH limits), which also restrict the allowed values of $f_{\rm PBH}$ shown above. The darker blue shading extends beyond the bare PBH outflow line to indicate that no viable parameter space exists in the upper-right corner.  
[Right] Comparison of interstellar gas heating bounds for dPBHs that include dark halos (lower black curve, hatched up to the corresponding bare PBH bound shown in the upper black curve) against existing constraints on PBHs. Above the thin blue dotted line at $f_{\rm allowed}=10^{-2}$, the dPBH bound becomes unreliable since halo formation could be disrupted. Shown for reference are existing bounds on PBHs without halos, which might be affected by the presence of dark halos: caustic-crossing microlensing in Icarus~\cite{2018PhRvD..97b3518O} (purple, I), gravitational wave O3 observations by LIGO-Virgo-KAGRA~\cite{LIGOScientific:2019kan,Andres-Carcasona:2024wqk} (gray, GW) - analysis of which assumed a mixed PBH and astrophysical black hole population with a log-normal PBH mass function, CMB constraints on accretion from Planck~\cite{Ali-Haimoud:2016mbv,Serpico:2020ehh} (orange, P), X-ray binary counts~\cite{Inoue:2017csr} (green, XRB), dynamical friction of halo objects~\cite{Carr:2018rid} (red, DF), dynamical friction effects on stellar half-light radii~\cite{Wadekar:2022ymq} (light blue, Star), Lyman-$\alpha$ structure bounds~\cite{2019PhRvL.123g1102M,Ivanov:2025pbu} (maroon, Ly-$\alpha$), survival of astrophysical systems in Eridanus II~\cite{Brandt:2016aco}, Segue 1~\cite{2017PhRvL.119d1102K}, and wide binary disruption~\cite{2014ApJ...790..159M} (magenta, S), large scale structure considerations~\cite{Carr:2018rid} (cyan, LSS), and isocurvature perturbations from the CMB~\cite{Gerlach:2025vco} (dark orange, CMB). These external bounds apply to bare PBHs and should be appropriately modified for the case of dressed PBHs.
} 
\label{fig:falloweddhcompare}
\end{figure} 

Since dPBH system contains $\mathcal{O}(100)$ times more mass than a bare PBH, the dynamical friction force of Eq.~\eqref{eq:FDFparameterize} is enhanced approximately by the square of this factor. For supersonic motion with $\mathcal{M}>1$, the Coulomb logarithm in Eq.~\eqref{eq:IMachpointmass} is reduced by $\sim 30$ due to the larger effective radius $r_{\rm min}$ of the DM halo compared to the central black hole itself. However, this suppression is more than compensated by the increased mass leading to significantly stronger dynamical friction effects and hence resulting bounds for dPBHs, as shown in the left panel of Fig.~\ref{fig:falloweddhcompare}.
  
The extended dPBH halo also modifies the accretion flow. Comparing the Bondi radius of  Eq.~\eqref{eq:rbondi} with the halo radius of Eq.~\eqref{eq:rh}, one finds $R_{\rm Bondi} < R_h$ for $M_{\rm PBH}\lesssim 10^{3} M_\odot$. Thus, only the enclosed mass of a dPBH within a Bondi radius affects gas accretion, which can be expressed as an integral using Eq.~\eqref{eq:Mh}, Eq.~\eqref{eq:rhoh}, and Eq.~\eqref{eq:rh},
\begin{equation}
\label{eq:menclosed}
    M_{\rm enc} = M_{\rm PBH} + \dfrac{4\pi}{3} \int_{0}^{R_{\rm Bondi}} \rho_h r^2 dr~ = M_{\rm PBH} + 4.29~M_\odot  \left(\frac{R_h}{\textrm{1~pc}}\right)^{9/4} \left(\frac{R_{\rm Bondi}}{\textrm{1~pc}}\right)^{3/4}~.
\end{equation}
The Bondi radius is then defined by Eq.~\eqref{eq:rbondi} as 
\begin{equation}
\label{eq:Bondimod}
     R_{\rm Bondi}  = ~ 1\textrm{ pc}\left(\frac{\tilde{v}}{10\textrm{ km/s}}\right)^{-2}\Big[8.60\times10^{-5} \left(\frac{M_{\rm PBH}}{  M_\odot}\right)  
      +3.77\times10^{-3} \left(\frac{M_{\rm PBH}}{ M_\odot}\right)^{3/4}\left(\frac{R_{\rm Bondi}}{1\textrm{ pc}}\right)^{3/4}\Big]~,  
\end{equation}
where the first term in parentheses corresponds to the PBH contribution and the second to the halo. The value of $R_{\rm Bondi}$ is then used to compute the enclosed mass $M_{\rm enc}$. 
The Bondi radius $R_{\rm Bondi}$ is then used to evaluate the enclosed mass $M_{\rm enc}$. For small PBH masses, the enclosed mass is only slightly larger than the bare mass with $M_{\rm enc}\gtrsim M_{\rm PBH}$. However, for $M_{\rm PBH}\gtrsim 10^3~M_\odot$, the enclosed mass asymptotes to include the contributions from the total halo mass with $M_{\rm enc}\to 10^2~M_{\rm PBH}$. Since the characteristic accretion disk radius  
$r_{\rm disk}\sim \mathcal{O}(10^3)R_s$~\cite{Mahadevan:1996jf,Pringle:1981ds} is much smaller than both $R_{\rm Bondi}$ and the halo radius $R_h$, the internal disk dynamics are not significantly affected by the surrounding halo. Accordingly, we employ the semi-analytical disk solutions of Sec.~\ref{sec:disk} with $M=M_{\rm PBH}$  and take the minimum radius $r_{\rm min} = 3 R_s$  at the ISCO radius of bare PBHs, which we assume also to be the ISCO radius of dPBHs, while rescaling the accretion rate by the larger enclosed mass of Eq.~\eqref{eq:bondihoyle}.

In the left panel of Fig.~\ref{fig:falloweddhcompare}, we present the gas heating constraints on dPBHs derived in this work, showing the individual contributions from dynamical friction, photon emission and winds, as well as the combined bounds. 
The shaded regions illustrate the difference between the stronger dPBH limits and the weaker bare PBH limits~\cite{Lu:2020bmd,Takhistov:2021aqx}, with hatching denoting difference between the combined constraints on bare PBHs and dPBHs.   
For $M_{\rm PBH}\gtrsim 10^{2}M_\odot$, the enhanced accretion rates for dPBHs become the dominant effect as $R_h \gtrsim R_{\rm Bondi}$. In contrast, for $M_{\rm PBH}\lesssim 10^{2}M_\odot$, the dominant improvement arises from the increased dynamical friction force, which is independent of the accretion rate.

The right panel of Fig.~\ref{fig:falloweddhcompare} compares the dPBH bounds, expressed in terms of the bare PBH mass $M_{\rm PBH}$, together with the bare PBH gas-heating constraints derived in Refs.~\cite{Lu:2020bmd,Takhistov:2021aqx}. For reference, we also overlay existing limits on (bare) PBHs, obtained without accounting for DM halos or their effects. Since  halo masses are typically $\mathcal{O}(100)$ times larger than seeding PBH masses, dressed halos cannot fully develop if PBHs constitute a significant DM fraction. In Fig.~\ref{fig:falloweddhcompare}, the blue dot-dashed line marks the threshold $f_{\rm PBH}\gtrsim 10^{-2}$  above which the ambient DM reservoir is insufficient to build full halos. A detailed treatment of these effects is left for future work.

\subsection{Axion minicluster subhalos}

The DM halo may contain abundant substructure in the form of self-virialized sub-halo clumps. In the case of axions\footnote{Here and in the following we collectively refer to axions and axion-like particles as axions.}, such substructures arise naturally if the Peccei–Quinn symmetry is broken after inflation. Primordial fluctuations in the axion field then evolve non-linearly after matter–radiation equality and can collapse into bound axion miniclusters. N-body simulations indicate~\cite{Xiao:2021nkb} that axion miniclusters are well described by NFW density profiles~\cite{Navarro:1996gj}  
\begin{equation}
    \rho(r) = \dfrac{\rho_s}{\dfrac{r}{r_{s}}\left(1+\dfrac{r}{r_{s}}\right)^2}~, \label{eq:NFWamc}
\end{equation}
where $\rho_s$ and $r_{\rm s}$ denote the characteristic density and scale radius, respectively. 

The effective radius of axion miniclusters can be described by the virial radius $r_{\rm vir}$,  conventionally defined to encompass an average overdensity of $\sim200$ relative to the background. The concentration parameter $\mathcal{C} = r_{\rm vir}/r_{\rm s}$ determines the central density enhancement, with simulations typically finding $\mathcal{C} \sim \mathcal{O}(10^2)$ for axion miniclusters~\cite{Xiao:2021nkb}. The virial radius $r_{\rm vir}$ is set by the enclosed mass and background cosmology. For an NFW profile, the virial mass of an axion minicluster  is   
\begin{equation}
     M_{\rm vir} = \int_0^{r_{\rm vir}} 4\pi r^2 \rho(r) dr = 4\pi  r_{s}^3 ~\rho_s\left(\ln (1+\mathcal{C}) - \frac{\mathcal{C}}{1+\mathcal{C}} \right)~. \label{eq:Mamc}
\end{equation}
Here, for EDCOs we take $M = M_{\rm vir}$.

At formation, around matter–radiation equality, the resulting characteristic minicluster mass is~\cite{Chang:2024fol}
\begin{equation}
    M_{\rm vir,c} \simeq 1.4\times 10^{4} ~M_\odot ~ \left(\frac{m_a}{10^{-18}   ~\text{eV}} \right)^{-3/2} \left(\frac{g_*(T_{\rm osc})}{g_*(T_0)} \right)^{1/4} \label{eq:Mamcc}
\end{equation}
with $g_*(T)$ being the entropic degrees of freedom at temperature $T$, evaluated at the axion oscillation temperature $T_{\rm osc}$ and the present CMB temperature $T_0$. We focus on light axions, $m_a\lesssim10^{-15}$ eV, for which $M_{\rm vir,c}$ lies in the mass range of interest. The mass of an individual minicluster can be expressed as $M_{\rm vir} = x_{\rm amc} M_{\rm vir,c}$ with $1 \lesssim x_{\rm amc}\lesssim 100$. For $x_{\rm amc}\lesssim 50$, denser core ``axion stars'' can form at minicluster centers~\cite{Chang:2024fol}, as we discuss below. 

The axion minicluster scale radius is approximately~\cite{Chang:2024fol}
\begin{equation}
    r_{s} \simeq   10^{12}   ~\text{km} ~ x_{\rm amc}^{5/6}   \left(\frac{M_{\rm vir,c}}{10^{4}   M_\odot}\right)^{1/3}. \label{eq:rs}
\end{equation} 
With concentration at formation given by $\mathcal{C} \simeq 4.1   x_{\rm amc}^{-1/2}$,   the characteristic density is 
\begin{equation}
    \rho_s \simeq 8.9 \times 10^8~   \text{GeV}~ \text{cm}^{-3}~ x_{\rm amc}^{-3/2}   \left[\ln (1+ 4.1   x_{\rm amc}^{-1/2}) - \dfrac{4.1   x_{\rm amc}^{-1/2}}{1+4.1   x_{\rm amc}^{-1/2}} \right]^{-1}. \label{eq:rhos}
\end{equation} 
After formation, the internal density profile remains fixed, while the average  
density of the Universe  
decreases. Consequently, the virial radius and hence the effective concentration grows with time, with simulations giving~\cite{Xiao:2021nkb}
\begin{equation}
    \mathcal{C}\simeq 1.4\times 10^4~   (1+z)^{-1} x_{\rm amc}^{-1/2}~. \label{eq:c}
\end{equation}

Combining Eqs.~\eqref{eq:NFWamc}–\eqref{eq:rhos}, the present-day virial radius is
\begin{equation}
     r_{\rm vir} = 1.0\times 10^{12} ~ \text{km} ~\mathcal{C}   x_{\rm amc}^{1/2}   \left(\dfrac{M_{\rm vir}}{10^{4}   M_\odot}\right)^{1/3}   \left(\dfrac{\ln (1+ 4.1   x_{\rm amc}^{-1/2}) - \dfrac{4.1   x_{\rm amc}^{-1/2}}{1+4.1   x_{\rm amc}^{-1/2}}}{\ln (1+ \mathcal{C}) - \dfrac{\mathcal{C}}{1+\mathcal{C}}}\right)^{1/3}~. \label{eq:Ramc}
\end{equation}

The outer envelopes of axion miniclusters are susceptible to tidal stripping and disruption within galactic halos~\cite{spitzer1958disruption, Green:2006hh}, although their dense cores are expected to survive over a Hubble time. In a dwarf galaxy environment such as Leo T where the stellar density is low, disruptive encounters are reduced~\cite{Xiao:2021nkb} and the dominant effect is truncation of the outer profile once the minicluster density falls below the ambient DM halo density. Requiring the density at the effective virial radius to equal the mean DM density of Leo T implies an effective present-day concentration $\mathcal{C}\simeq200$ for $x_{\rm amc}=100$.

\begin{figure}
    \centering
    \includegraphics[width=1\linewidth]{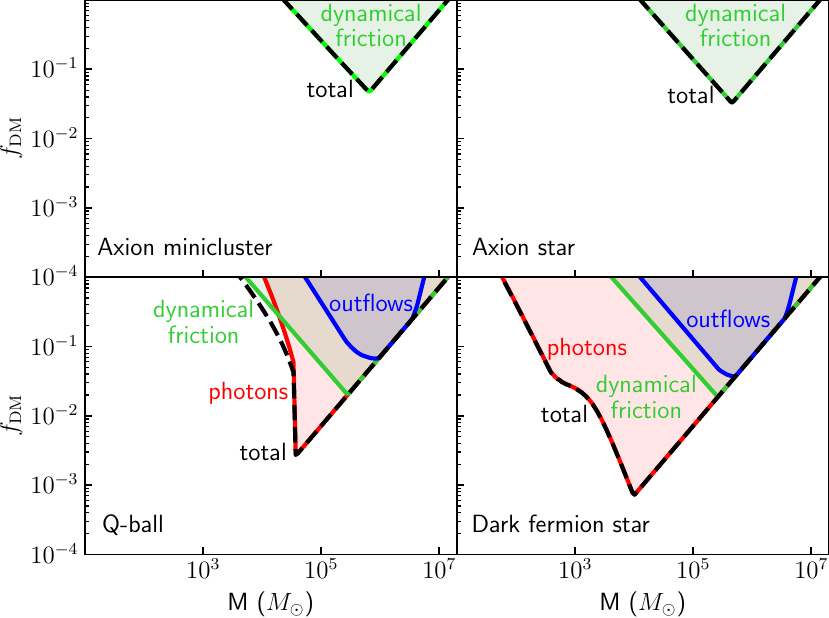}
    \caption{Interstellar gas heating constraints derived in this work on the DM fraction in EDCOs including axion miniclusters [Top left], axion stars [Top right], Q-balls [Bottom left], and dark fermion stars [Bottom right], considering reference model parameters from Sec.~\ref{sec:selected}. Contributions from different heating channels are shown: accretion-disk photons (red), dynamical friction (green), and outflows (blue), with combined bounds given by black dashed lines. In contrast to the dPBH case, the excluded parameter regions are shown upward to $f_{\rm DM} = 1$.}
    \label{fig:selectedbounds}
\end{figure}

In Fig.~\ref{fig:compactranges} we illustrate the mass–radius relation for $10\lesssim x_{\rm amc}\lesssim100$ and $200\lesssim \mathcal{C}\lesssim4400$, highlighting the reference model we consider with the violet curve. In Fig.~\ref{fig:selectedbounds} we show the resulting gas heating bounds for axion miniclusters derived in this work, adopting $x_{\rm amc}=100$ and $\mathcal{C}=200$. The corresponding mass range $10<M_{\rm vir}/M_\odot<10^7$ maps to axion masses  $2.7\times10^{-19}~\text{eV}~\lesssim m_a \lesssim 2.7\times10^{-15}~\text{eV}$ considering Eq.~\eqref{eq:Mamcc}.  

\subsection{Axion stars}

After miniclusters form in the early Universe, they can undergo mergers and tidal interactions in galactic halos and subsequently seed the formation of gravitationally bound solitonic axion stars~\cite{Kolb:1993zz,Levkov:2018kau,Eggemeier:2019jsu,Kirkpatrick:2020fwd}. The density profile of axion stars indicated by numerical simulations is $\rho(r) \propto (1+0.091(r/r_{\rm hm})^2)^{-8}$, for a half-mass radius $r_{\rm hm}$~\cite{Schive:2014dra, Schive:2014hza, Ellis:2022grh}. This form is well approximated by a Gaussian profile~\cite{Chavanis:2011zi, Chavanis:2011zm, Chavanis:2016dab, Visinelli:2017ooc, Chang:2024fol}
\begin{equation}
    \rho(r) = \rho_0   e^{-r^2 / r_*^2}~. \label{eq:Gaussianas}
\end{equation} 
The axion star radius can then be defined as the radius enclosing 90\% of the total energy. For a Gaussian distribution, this yields~\cite{Visinelli:2017ooc}   $R= 1.77 r_*$.

The mass $M$ and radius $R$ of axion stars depend on the axion mass and the strength of self-interactions. The equilibrium configuration is obtained by minimizing the total energy, leading to two branches of solutions.
One is a dilute branch, in which gravity is balanced by quantum (gradient) pressure. The other is a dense branch, in which attractive self-interactions dominate over gravity. The dilute branch is stable and long-lived, while the dense branch is unstable to collapse. In the absence of significant self-interactions, only the stable dilute branch exists, with radius~\cite{Chang:2024fol}
\begin{equation}
    R (M) \simeq 2.6 \times 10^{13}~\text{km}~\left(\frac{m_a}{10^{-18}~   \text{eV}} \right)^{-2} \left(\frac{M}{10^{4}~M_\odot} \right)^{-1}~. \label{eq:Ras}
\end{equation}
If self-interactions are non-negligible, the dilute branch terminates at a maximum mass and continued growth can trigger a collapse instability resulting in ``bosenova'' emission of relativistic axions with distinct signatures~\cite{Eby:2016cnq,Levkov:2016rkk,Eby:2021ece, Arakawa:2023gyq,Eby:2024mhd,Arakawa:2024lqr,Arakawa:2025hcn}. On the other hand, dense axion stars can be stabilized by higher order repulsive interactions~\cite{Eby:2016cnq}.

In the weak self-interaction regime, the characteristic mass scale of axion stars formed via gravitational accumulation is~\cite{Chang:2024fol} 
\begin{equation}
    M \simeq 3.7 \times 10^{4}~M_\odot  ~x_{\rm amc}^{1/12}   \left(\frac{m_a}{10^{-18}~\text{eV}} \right)^{-3/2} \left(\frac{g_*(T_{\rm osc})}{g_*(T_0)} \right)^{1/12}~,~\label{eq:Mas}
\end{equation}
with growth slowing above this scale. For miniclusters with $x_{\rm amc} \sim \mathcal{O}(1)$, nearly all axions can be incorporated into axion stars. The fraction of axions bound in stars decreases with increasing $x_{\rm amc}$, vanishing for up to $x_{\rm amc} \gtrsim 50$.
 
In the top-right panel of Fig.~\ref{fig:selectedbounds}, we present the total gas heating bounds on axion stars derived in this work, fixing $x_{\rm amc}=1$. For this reference model, the axion star mass range  
$10<M/M_\odot<10^7  $ corresponds to  
the axion mass range $2.4\times10^{-20}\textrm{ eV} \lesssim m_a \lesssim 2.4\times10^{-16}\textrm{ eV}$ through Eq.~\eqref{eq:Mas}. In Fig.~\ref{fig:compactranges}, the reference model is indicated with the solid blue line together with the mass–radius range corresponding to $1<x_{\rm amc}<50$.

\subsection{Q-balls}

Q-balls are non-topological solitons that can be abundantly produced in the early Universe from instabilities of interacting charged scalars~\cite{Coleman:1985ki}. Their detailed properties depend on the underlying scalar potential and can lead to a variety of phenomenological consequences\footnote{This includes possible formation of PBHs~\cite{Cotner:2016cvr,Cotner:2019ykd}.}, including DM ~\cite{Kusenko:1997si, Enqvist:1997si}. Among the different Q-ball realizations, we focus here on the single-field sextic (Coleman) Q-ball with interactions up to $\phi^6$ as reference~(e.g.~\cite{Heeck:2020bau,Ansari:2023cay}). For simplicity, we work in the thin-wall and large-$Q$ limit, assuming Q-balls interact only gravitationally with SM particles and are effectively stable, do not decay or collapse. Other prominent Q-ball realizations motivated by supersymmetric theories, such as gauge-mediated and gravity-mediated Q-balls, feature different mass–radius relations~\cite{Kusenko:2005du,Loiko:2022noq}.

We consider Lagrangian containing a single complex scalar $\Phi$ of mass $m$ with interactions up to $\phi^6$ 
\begin{equation}
    \mathcal{L} = \partial_\mu \Phi\partial^\mu \Phi^\dagger - m^2\phi^2 -\lambda \phi^4 - \zeta\phi^6  ,
\end{equation}
with $\Phi = e^{i\omega t}\phi$. 
Here, $\omega$ is the angular frequency and also the Lagrange multiplier which determines the conserved charge $Q$  
\begin{equation}
    Q = i\int d^3 r \Phi \partial_0 \Phi^\dagger - \Phi^\dagger \partial_0 \Phi = 2\omega \int d^3 r~ \phi~.
\end{equation}

In the thin-wall limit $\omega\xrightarrow{} \omega_Q$, with Q-ball mass 
\begin{equation} \label{eq:qballmass}
M=\omega_Q Q
\end{equation}
and
\begin{equation}
    \omega_Q = m\sqrt{1-\frac{\lambda^2}{4m^2\zeta}}~.
\end{equation}
Considering $\theta = \sqrt{m^2-\omega_Q^2} = \sqrt{\lambda^2/4\zeta}$ and $\phi_Q = \sqrt{|\lambda|/2\zeta}$ as functions of the coupling constants, the radius $R_Q$ is given by
\begin{equation}
    R_Q = \left(\frac{3\omega}{4\pi\omega_Q\phi_Q^2}\int_0^{\infty}d^3 r\phi(r)^2\right)^{1/3}
    = \left(\frac{3Q}{8\pi \omega_Q \phi_Q^2}\right)^{1/3}~.
    \end{equation}
The density profile of thin-wall Q-balls is approximately uniform 
\begin{equation}
    \rho(r) = 2\phi_Q^2 \omega_Q^2 \Theta(R_Q-r)~.\label{eq:Qballdensity}
\end{equation}  

In Fig.~\ref{fig:selectedbounds} we display gas heating bounds derived in this work for Coleman-type Q-balls,  using as the fiducial model $m=50~\mathrm{keV}$, $\lambda=-1$, $\zeta=1/m^2$. 
The resulting Q-ball radius can be comparable to the ISCO of a PBH at larger mass of $M\sim10^7~M_\odot$, but is relatively diffuse at lower masses. The Q-ball mass and radius ranges shown in Fig.~\ref{fig:compactranges} correspond to the parameter ranges $\mathrm{keV}\lesssim m \lesssim \mathrm{MeV}$ and $-1\lesssim \lambda \lesssim -0.1$, with $\zeta=1/m^2$. We restrict to parameters for which the minimum radius satisfies $r_{\rm min} > 3 R_s$, ensuring that no horizon or ISCO forms. Near this limit, the flat-spacetime approximation~\cite{Ansari:2023cay} begins to break down.  In contrast to axion stars or miniclusters, Q-ball properties are less tightly constrained by microphysics, hence we adopt representative benchmark values motivated by the sextic Coleman potential. For the Q-ball mass range $10<M/M_{\odot}<10^7$ in Fig.~\ref{fig:selectedbounds}, from relation of Eq.~\eqref{eq:qballmass} the conserved charge spans the range $10^{62}\lesssim Q\lesssim10^{68}$ and this determines the configuration compactness.

Beyond Coleman Q-balls, other well-studied classes of Q-balls exist. 
In SUSY theories with gauge-mediated breaking~\cite{Kusenko:1997si, Enqvist:1997si}, nearly flat scalar potentials 
of the form 
\begin{equation}
    V(\phi) \sim M_S^4 ,
\end{equation}
where $M_S$ is the soft SUSY-breaking scale, give rise to gauge-mediated Q-balls with
\begin{equation}
    M \propto M_S Q^{3/4}, 
    \qquad 
    R \propto M_S^{-1} Q^{1/4}.
\end{equation}
These objects are diffuse, with much lower compactness than Coleman thin-wall Q-balls of the same mass, 
and therefore yield weaker gas-heating bounds. 
In contrast, in gravity-mediated SUSY breaking~\cite{Kasuya:2000wx}, the scalar potential takes the form 
\begin{equation}
    V(\phi) \sim m_s^2\left(1+K\ln \left(\frac{\phi^2}{M^2}\right)\right)\phi^2 ,
\end{equation}
with soft scalar mass $m_s$ and loop correction parameter $K<0$. 
The resulting gravity-mediated Q-balls satisfy
\begin{equation}
    M \simeq m_s Q, 
    \qquad 
    R \sim |K|^{-1/2} m_s^{-1},
\end{equation}
essentially independent of $Q$. 
These are more compact and behave closer to point-mass objects, leading to stronger heating constraints.

 We emphasize that in this work we treat Q-balls as transparent to baryons and interacting only gravitationally, so that their heating arises solely from gravitational wakes and potential accretion flows. In contrast, other EDCOs such as dark quark nuggets\footnote{See Ref.~\cite{Witten:1984rs,Madsen:1998uh} for purely visible sector quark nuggets.}~\cite{Bai:2018dxf} could be opaque to baryons and then behave effectively as hard spheres with nuclear densities. Their gas heating signatures can therefore be modeled within our framework as highly compact uniform-density objects, closely resembling the thin-wall transparent Q-ball scenario but with possible strong non-gravitational interactions.

\subsection{Dark fermion stars}

Dark fermion stars are highly compact self-gravitating objects, analogous to neutron stars but composed of heavy DM fermions. Unlike axion stars, which are diffuse due to the light axion masses, dark fermion stars can approach black hole compactness. Several related EDCOs have been also explored in the literature including dark exoplanets~\cite{Bai:2023mfi} and mirror stars~\cite{Kobzarev:1966qya,Hippert:2021fch}. 
 
We adopt the mass–radius relations of Ref.~\cite{Narain:2006kx,Gresham:2018rqo}, obtained by solving the Tolman - Oppenheimer - Volkoff equations for asymmetric fermion matter. Taking the extremal stable branch, one finds for repulsive interactions
\begin{equation}
\label{eq:Mdfs}
    M = 1.63~M_\odot~(0.384 + 0.165y)\left(\frac{m_f}{1~\textrm{GeV}}\right)^{-2}
\end{equation}
and
\begin{equation}
\label{eq:Rdfs}
    R = 2.41 \textrm{ km}~(3.367+0.797y)\left(\frac{m_f}{1~\textrm{GeV}}\right)^{-2}  ~,
\end{equation}
where $m_f$ is the fermion mass and $y = m_f/m_I$ is the ratio to the interaction mass scale. These scalings are valid both in the strong-interaction limit $y\gg1$, such as for neutralino-like interactions, and in the weak-interaction regime $y\ll1$. If the dark fermion is thermally produced, Big Bang nucleosynthesis   requires $m_f\gtrsim10~\mathrm{MeV}$~\cite{Balan:2024cmq}. The density profiles found in~Ref.~\cite{Narain:2006kx} are well approximated by a Gaussian, allowing us to use the same framework developed for axion stars.

In Fig.~\ref{fig:selectedbounds}, we display gas heating constraints derived in this work for dark fermion stars.
Due to their extreme compactness, dark fermion stars can efficiently form accretion disks. Heating of the interstellar medium then arises both from dynamical friction and from radiative emission, photons and outflows, associated with the accretion process. In the reference benchmark shown in   Fig.~\ref{fig:selectedbounds}, we consider $y=5\times10^{3}$, corresponding to dark fermion masses in the range $10~\mathrm{MeV}\lesssim m_f \lesssim 10~\mathrm{GeV}$. For the variation of the radius in the red shaded band in Fig.~\ref{fig:compactranges}, we scan across the parameter  $10^{-2}<y<5\times10^{3}$~\cite{Narain:2006kx} within the same fermion mass interval.  

\subsection{Comparison of constraints for extended compact objects}

\begin{figure}[t]
\centering
\includegraphics[width=0.6\linewidth]{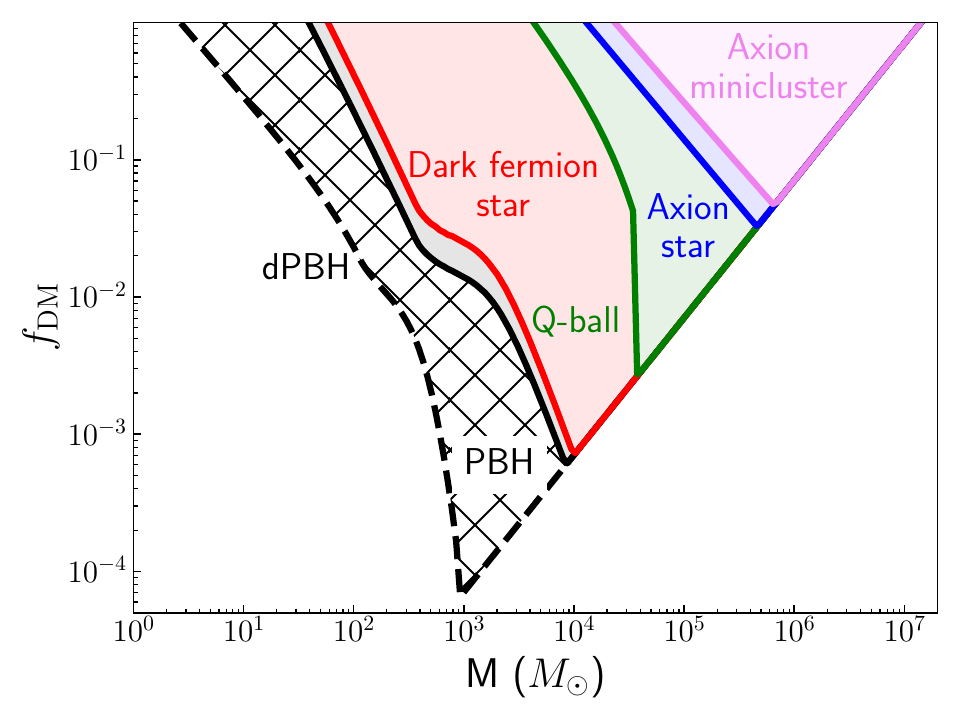}
\caption{
Combined interstellar gas heating constraints derived in this work on the DM fraction in EDCOs including dPBHs (dashed black), dark fermion stars (red), Q-balls (green), axion stars (blue), and axion miniclusters (violet). For comparison, gas-heating limits on bare PBHs without DM halos are shown as the solid black curve. The limits correspond to representative model parameters defined in Sec.~\ref{sec:selected}, with fiducial mass–radius relations indicated by solid lines. For dPBHs, the constraints are expressed in terms of the central PBH mass $M_{\rm PBH}$ and associated DM fraction $f_{\rm PBH}$.} 
\label{fig:fallowedcompareleotco}
\end{figure}

In this section, we analyzed gas heating for several distinct classes of EDCOs - dPBHs, axion miniclusters, axion stars, Q-balls, and dark fermion stars. We further compared the results to the case of bare PBHs without halos, which serve as the standard reference for compact DM phenomenological effects.

dPBHs are treated distinctly since they are composite systems consisting of a central PBH surrounded by a DM halo of mass $\mathcal{O}(100)$ times larger. In Fig.~\ref{fig:falloweddhcompare} and~Fig.~\ref{fig:fallowedcompareleotco}, the dPBH constraints are expressed in terms of the bare PBH mass, excluding the halo contribution. Only the fraction of the halo mass enclosed within the Bondi radius contributes to accretion, thus the effective accreting mass is greater than the central PBH mass~as described by Eq.~\eqref{eq:menclosed}. The resulting dPBH bounds in Fig.~\ref{fig:fallowedcompareleotco} are therefore stronger than those for bare PBHs, reflecting the $\dot{M}\propto M_{\rm enc}^2$ scaling of the accretion rate. A further enhancement arises because the gas is captured at the Bondi radius but disk formation and radiative processes occur much closer to the black hole, where the gravitational potential is deeper. Consequently, the dimensionless accretion rate $\dot{m}$ that is defined as the accretion rate by the enclosed mass normalized to the Eddington rate of the central PBH more readily reaches the thin- and slim-disk regimes, which are characterized by higher radiative efficiency and higher disk temperatures.

The analyses of four other EDCO classes - dark fermion stars, Q-balls, axion stars, and axion miniclusters - in the order of increasing radius illustrate a key trend that more compact objects generically yield stronger gas heating bounds. This arises because heating efficiencies scale directly with the innermost radius $r_{\rm min}$ that sets the gravitational potential felt by the accreting gas. At smaller $r_{\rm min}$, the potential is deeper, enhancing photon emission and outflows. For diffuse EDCOs with large $r_{\rm min}$, these channels are suppressed and dynamical friction effects dominate. Since dynamical friction enters logarithmically through $\ln(r_{\rm sys}/r_{\rm min})$ in Eq.~\eqref{eq:IMachpointmass}, the suppression is mild but sufficient to weaken bounds for more diffuse objects. Outflows typically remain subdominant in our modeling, considering weak winds as discussed in Sec.~\ref{ssec:thick}, although for efficient winds or jet formation these effects can be significantly enhanced~\cite{Takhistov:2021aqx,Takhistov:2021upb}. At higher EDCO masses $M\gtrsim10^6~M_\odot$, Q-balls, dark fermion stars, and bare PBHs transition from thick-disk to thin-disk accretion regimes, at which point winds disappear entirely. In contrast, dPBHs can sustain slim-disk accretion with powerful winds even at high masses with a nonvanishing outflow contribution.

As shown in Fig.~\ref{fig:fallowedvary}, generalized EDCOs with $R=3R_s$ and $R=10R_s$ are dominated by photon heating, whereas by $R=100R_s$ dynamical friction dominates until the thin-disk regime is reached around $M\sim 5\times10^4~M_\odot$. The reference dark fermion star model with radius $R\simeq 4R_s$  is nearly as compact as a Schwarzschild PBH that has ISCO at $R=3R_s$ and thus lies in the regime dominated by photon heating. In contrast, Q-balls are more diffuse and dynamical friction dominates their bounds until the thin-disk transition where photon emission becomes significant, producing the characteristic peak in Fig.~\ref{fig:selectedbounds}. For axion stars, the radius is sufficiently large that neither photon emission nor outflows are efficient. Their gas heating bounds are thus dominated by dynamical friction effects, and for sufficiently small masses, their diffuseness prevents even formation of thin disk as described by Eq.~\eqref{eq:thindiskcond}. Finally, the weaker constraints on axion miniclusters relative to axion stars arise from their larger effective radii entering the Coulomb logarithm of the dynamical friction expression, which reduces the overall drag force. This is only partially compensated by  larger finite-size corrections $I_{\rm ext}$ of NFW profiles compared to Gaussian profiles as show in Tab.~\ref{tab:Reff}, which enhance the drag by accounting for extended mass distributions.  

We thus find that bounds from dynamical friction are shaped by two competing finite-size effects. One is related to larger effective radii entering the Coulomb logarithm, which suppress the drag force. The other is related to the profile-dependent finite-size correction $I_{\rm ext}$, which enhances the wake in the supersonic regime. For compact objects such as Q-balls or Gaussian-profile axion and fermion stars, $I_{\rm ext}$ is modest and the constraints scale primarily with compactness. On the other hand, for highly concentrated EDCOs such as NFW profiles as in axion miniclusters $I_{\rm ext}$ can be significant, partially offsetting the suppression from larger radii. This competition between the Coulomb logarithm and $I_{\rm ext}$ represents one of the new results of our analysis and highlights the importance of consistently treating extended mass profiles rather than relying on effective point-mass prescriptions.

\section{Conclusions}
\label{sec:conclusions}

We have analyzed interstellar gas heating from a broad range of EDCOs including dPBHs, axion miniclusters, axion stars, Q-balls, dark fermion stars, as well as more generalized configurations. Unlike ordinary astrophysical bodies, these objects are effectively transparent to interstellar gas, which requires an extension of the standard dynamical friction formalism to incorporate finite size, internal density profiles  and gas penetration effects. We derived a new finite-size correction term $I_{\rm ext}$  that depends on the internal mass distribution of the EDCO and converges rapidly in time. This demonstrates that the near-field wake primarily determines the additional drag effects. Establishing the interplay between $I_{\rm ext}$ and the Coulomb logarithm represents a key new development of this work, broadly applicable across many theoretical models.

We further generalized accretion disk emission and outflows to horizonless EDCOs with $r_{\rm min}>R_s$, adapting semi-analytic models from black holes. We find that dPBHs accrete at higher rates than bare PBHs and can form slim disks. Such regimes occur only at very high masses and do not qualitatively change the resulting gas-heating constraints, but give additional insights into the range of possible accretion behavior.

Considering the thermal balance of the Leo T dwarf galaxy, we combined gas heating effects from dynamical friction, photon emission, and outflows to derive new constraints across distinct EDCO classes. Surrounding halos of dPBHs substantially boost accretion rates, leading to bounds that are significantly stronger than for bare PBHs without halos. For horizonless EDCOs, compactness in the mass-radius relation plays a key role. Objects with radii only a few times $R_s$, such as dark fermion stars and Q-balls, can efficiently form accretion disks and are constrained primarily by photon emission. On the other hand, more diffuse configurations such as axion stars and miniclusters are limited by dynamical friction. In the latter case, the bounds are weaker but remain non-negligible due to the logarithmic sensitivity to $r_{\rm min}$ and can be 
enhanced by $I_{\rm ext}$ for concentrated profiles (e.g. NFW-like).

In summary, our study establishes the first unified gas heating constraints on EDCOs. The resulting limits depend sensitively on three key ingredients: (i) the presence of a surrounding dark halo, (ii) the compactness of the EDCO  and (iii) the internal density profile through the finite-size correction $I_{\rm ext}$ effects. Quantitatively, for dPBHs dark halos surrounding PBHs enhance accretion and gravitational drag, strengthening the resulting bounds by up to two orders of magnitude compared to bare PBHs. For horizonless EDCOs with $r_{\rm min} > 3R_s$, the photon emission efficiency decreases with compactness, so that extremely diffuse configurations such as axion stars and axion miniclusters are constrained primarily by the dynamical friction. Finally, the newly derived $I_{\rm ext}$ correction to dynamical friction can increase the heating rate and tighten constraints by up to a factor of $\sim 2$ for the large radius EDCOs and parameters considered here. Together, we find that these effects significantly modify the excluded parameter space, providing a general and robust framework for assessing interstellar gas heating bounds across diverse compact dark sector configurations, extending well beyond black holes and applicable to a wide variety of theoretical scenarios.

\section*{Acknowledgment}
\addcontentsline{toc}{section}{Acknowledgments}

This work was supported by World Premier International Research Center Initiative (WPI), MEXT, Japan. This work is supported by the Center for Advanced Computation at Korea Institute for Advanced Study. T.H.K. is supported by KIAS Individual Grant PG095202 at Korea Institute for Advanced Study. P.L. is supported by KIAS Individual Grant 6G097701. V.T. acknowledges support by the JSPS KAKENHI grant No. 23K13109.

\appendix

\section{Constraints on Generalized Compact Objects}

\begin{figure}[t]
\centering
\includegraphics[width=0.49\linewidth]{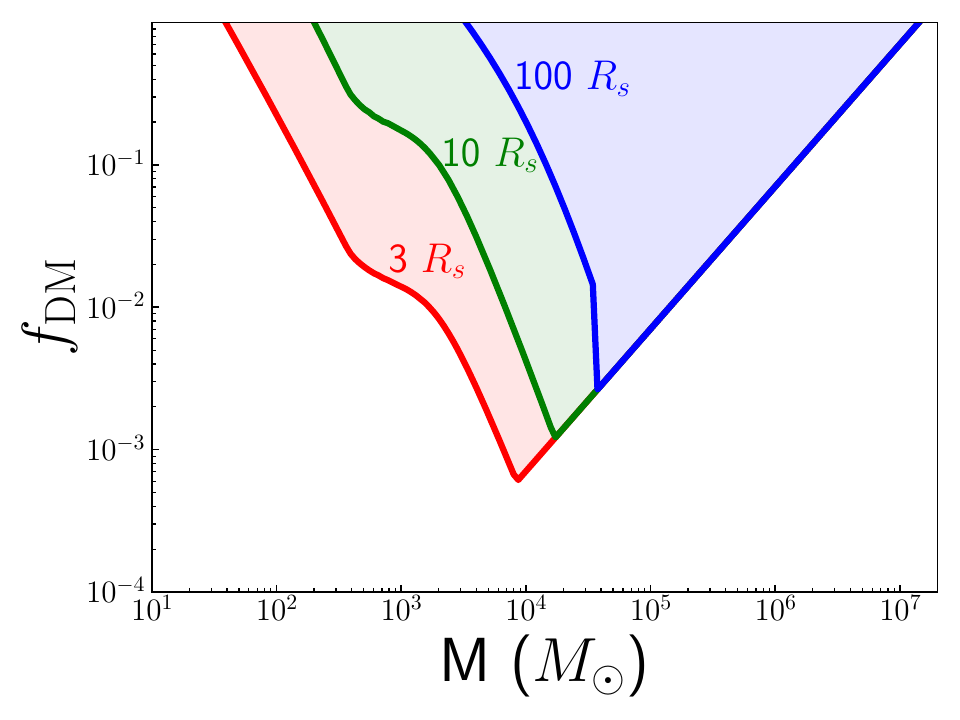}
\includegraphics[width=0.49\linewidth]{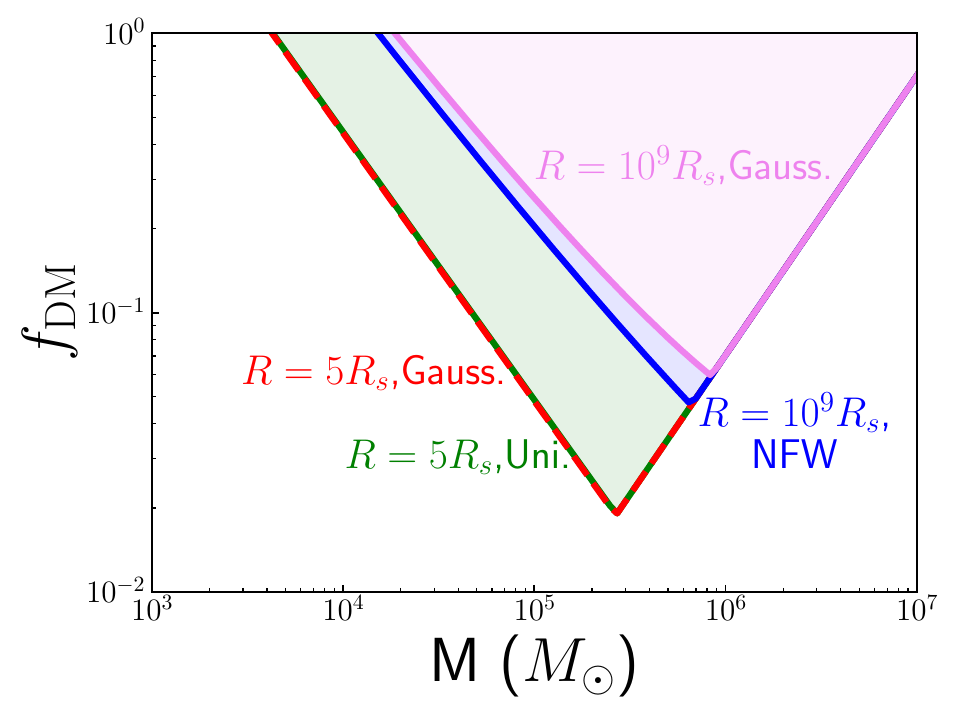}
\caption{[Left] Interstellar gas heating bounds on generalized EDCOs with PBH-like point-mass distributions and fixed radii $r_{\rm min} = R=3R_s$ (red), $R=10R_s$ (green), and $R=100R_s$ (blue). Constraints weaken with increasing radius, transitioning from photon-dominated heating at small radii to being dominated by dynamical friction heating at large radii. [Right] 
Interstellar gas heating bounds solely from dynamical friction, shown for $R=5R_s$ with a Gaussian profile (red), $R=5R_s$ with a uniform profile (green), 
$R=10^9R_s$ with an NFW profile (blue), and $R=10^9R_s$ with a Gaussian profile (violet). These benchmarks are representative of the dark fermion star, Q-ball, axion minicluster  and axion star models discussed in Sec.~\ref{sec:selected}, respectively. Although the radii are chosen in identical pairs of $5 R_s$ and $10^9 R_s$ here to highlight $I_{\rm ext}$ contributions, the reference Q-ball model typically has a larger radius than dark fermion stars  and axion miniclusters are generally larger than axion stars. Since the drag force depends only logarithmically on $r_{\rm min}$, the relative ordering of the dynamical-friction constraints for these EDCOs differs from that displayed in Fig.~\ref{fig:selectedbounds}. 
} 
\label{fig:fallowedvary}
\end{figure}

Here we extend the gas-heating analysis to generalized EDCOs characterized by fixed mass-radius ratios and a PBH-like mass distribution. Namely, we consider objects that behave as point masses but possess a finite physical radius. In the left panel of Fig.~\ref{fig:fallowedvary}, we show how the total gas heating constraints vary with the assumed minimum radius $r_{\rm min}$. For compact configurations with $r_{\rm min}=R=3R_s$ or $10R_s$, the dominant heating channel is photon emission from the accretion disk. By contrast, for more diffuse objects with $R=100R_s$ the constraints are set primarily by dynamical friction except for a peak around $M\sim 5\times10^4 M_\odot$, where photon emission becomes more prominent due to the transition into the thin-disk regime. This behavior reflects the much stronger dependence of photon luminosity on $r_{\rm min}$, as given by Eq.~\eqref{eq:Lnup} and~Eq.~\eqref{eq:thinspectrum}, compared to the weak logarithmic scaling of dynamical friction through the Coulomb logarithm in Eq.~\eqref{eq:IMachpointmass}. We observe that more compact objects with smaller radius induce more efficient gas heating and hence stronger resulting constraints.

The right panel of Fig.~\ref{fig:fallowedvary} showcases the isolated contributions from dynamical friction, highlighting sensitivity to EDCO radius variations and internal mass profiles. The benchmark parameters are chosen to resemble the reference axion minicluster, axion star, Q-ball, and dark fermion star models discussed in Sec.~\ref{sec:selected}. A large gap in radii, $5R_s$ compared to $10^9R_s$, produces an order of magnitude difference in the dynamical friction gas heating constraints through the Coulomb logarithm contributions. Differences in the finite-size correction $I_{\rm ext}$ between mass profiles are comparatively modest, particularly between uniform and Gaussian distributions shown in Tab.~\ref{tab:Reff}. A more noticeable effect arises for significantly extended objects with $R=10^9R_s$, where the relatively larger $I_{\rm ext}$ of NFW profiles compared to Gaussian profiles produces non-negligible differences in the gas heating bounds. Since the logarithmic ratio $\ln(r_{\rm sys}/r_{\rm min})$ decreases for increasing EDCO mass, the role of $I_{\rm ext}$ becomes more pronounced, leading to an increasing difference in the constraints
between profile models at high masses.

Our generalized analysis illustrates the same qualitative trends as observed for particular EDCO classes discussed in the main text. Namely, more compact objects typically yield stronger photon and outflow-related gas heating bound contributions, while dynamical-friction constraints depend sensitively on both the effective radius through the Coulomb logarithm and the internal mass profile through $I_{\rm ext}$.

\section{Parameter Dependence of Constraints}

\begin{figure}[t]
\centering
\includegraphics[width=0.49\linewidth]{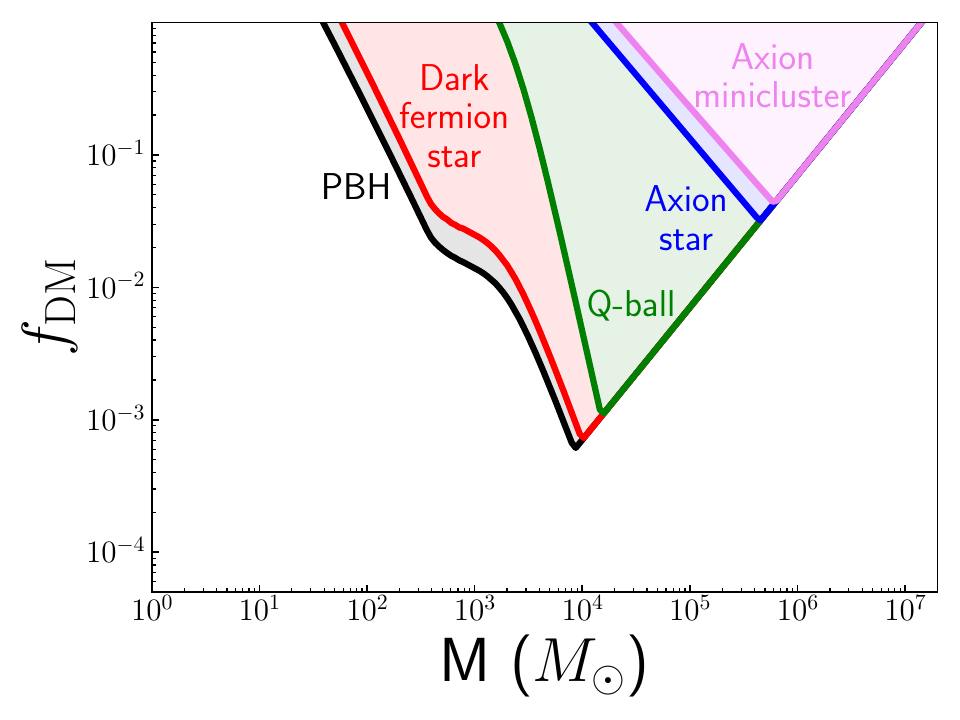}
\includegraphics[width=0.49\linewidth]{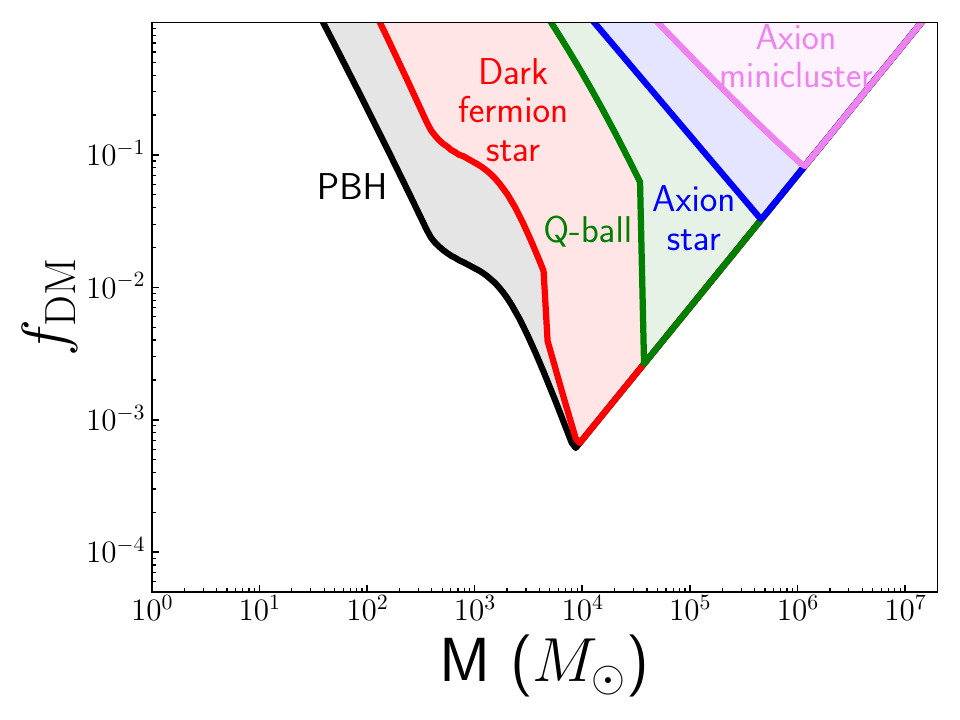}
\caption{[Left] Comparison of interstellar gas heating constraints for PBHs without DM halos (black), dark fermion stars (red), Q-balls (green), axion stars (blue), and axion miniclusters (violet), evaluated for the minimum radius models within the parameter ranges presented in Sec.~\ref{sec:selected} and shown in Fig.~\ref{fig:compactranges}. 
[Right]  Comparison of interstellar gas heating constraints for maximum radius models, which are more diffuse and therefore generally result in weaker bounds.
} 
\label{fig:fallowedminmax}
\end{figure}

In Sec.~\ref{sec:selected}, we considered reference benchmark parameters for each EDCO class with representative parameter ranges. These parameter ranges translate into a variety of EDCO mass-radius relations, as shown in Fig.~\ref{fig:compactranges}. To illustrate their impact on gas heating, we evaluate the resulting constraints considering minimal and maximal radii cases for each object type. In Fig.~\ref{fig:fallowedminmax}, we display the resulting bounds, with minimum radius models shown in the left panel and maximum radius models in the right panel. We do not display dPBHs as we do not consider here additional free parameters beyond the central PBH mass.

For axion miniclusters and axion stars, the bounds are determined primarily by dynamical friction. The broader variation of $f_{\rm DM}$ in the axion minicluster case arises primarily from their wider range of resulting potential radii as shown in Fig.~\ref{fig:compactranges}. Since the dynamical friction gas heating rate scales only logarithmically with EDCO radius, even orders of magnitude differences in size translate into relatively modest variations in the resulting constraints.  

Q-balls exhibit qualitatively different behavior between the minimal and maximal radii cases. In the minimum radius scenario, accretion disk photon emission becomes efficient already at $M\sim 10^3~M_\odot$, and the photon contributions to gas heating rapidly increase as the physical radius decreases in units of $R_s$. This results in a steep slope observed for the Q-ball constraints in the left panel of Fig.~\ref{fig:fallowedminmax}. In contrast, the maximum radius Q-ball scenario has significantly weaker photon emission such that
the gas heating bounds are dominated by dynamical friction until thin disk emission becomes prominent at $M\sim 3\times10^4 M_\odot$ as discussed in  Sec.~\ref{ssec:thin}.  We note that mass-radius relations are distinct for other types of Q-balls, such as gauge-mediated or gravity-mediated.

Dark fermion stars also show distinctive behavior, reflecting the structure of the mass–radius relations in Eq.~\eqref{eq:Mdfs} and~Eq.~\eqref{eq:Rdfs}. The minimum radius case coincides with the reference model with  $R\simeq 4R_s$, which is nearly as compact as a Schwarzschild black hole and hence the gas heating bound on $f_{\rm DM}$ here is only slightly weaker than for PBHs. In the maximum radius case, small fermion masses $m_f\simeq 10~\mathrm{MeV}$ and small interaction ratios $y\ll 1$ lead to radii $R\simeq 7R_s$ for low-mass objects. However, since $M$ increases with $y$ as in Eq.~\eqref{eq:Mdfs} and $m_f>10\textrm{ MeV}$,  large masses $M\gtrsim 5\times10^3~M_\odot$ require $y\gg 1$ so the maximum radius converges to the minimum radius $R\simeq 4R_s$ at high masses, producing the change of slope observed in Fig.~\ref{fig:fallowedminmax}.

\bibliography{refs}
\addcontentsline{toc}{section}{References}
\bibliographystyle{JHEP}

\end{document}